\documentclass{aa}  

\usepackage{natbib}
\bibpunct{(}{)}{;}{a}{}{,} 
\usepackage{graphicx}
\usepackage{tabularx}
\usepackage{supertabular}

\usepackage{threeparttable}
\usepackage{txfonts}
\usepackage{hyperref}

    \hypersetup{colorlinks=true,allcolors=[rgb]{0,0,1}}

\newcommand{\lya}{\mbox{Lyman-$\alpha$}\xspace}

\newcommand{\mm}{\mbox{$\,\mu\mathrm{m}$}\xspace}

\newcommand{\ewlya}{\mbox{$\mathrm{EW_{Ly\alpha}}$}\xspace}

\newcommand{\flya}{\mbox{$\mathrm{f_{esc,Ly\alpha}}$}\xspace}
\newcommand{\flyC}{\mbox{$\mathrm{f_{esc,LyC}}$}\xspace}
\newcommand{\rlya}{\mbox{$\rho_{\mathrm{Ly\alpha}}$}\xspace}
\newcommand{\rtot}{\mbox{$\rho_{\mathrm{total}}$}\xspace}

\defcitealias{GdlV2019LAELF}{dLV19}
\defcitealias{GdlV2020LAEfrac}{dLV20}
\defcitealias{richard2021atlas}{R21}
\defcitealias{AC2022LLAMAS}{C22}
\defcitealias{blanc2011}{B11}
\defcitealias{Thai2023}{T23}
\defcitealias{Rosdahl_2022}{R22}
\begin{document} 

   \title{Charting the \lya escape fraction in the range $2.9<z<6.7$ and consequences for the LAE reionisation contribution}
 \authorrunning{I. Goovaerts \& T. T. Thai} 
   \titlerunning{Charting the \lya escape fraction}

   \author{I. Goovaerts\inst{1}\and T. T. Thai\inst{1,2,3}\and R. Pello\inst{1}\and P. Tuan-Anh\inst{2,3} \and N. Laporte\inst{1} \and J. Matthee\inst{4} \and T. Nanayakkara\inst{5} \and J. Pharo\inst{6}}

   \institute{Aix Marseille Université, CNRS, CNES, LAM (Laboratoire d’Astrophysique de Marseille), UMR 7326, 13388 Marseille, France\\
              \email{ilias.goovaerts@lam.fr; tranthai.physics@gmail.com}
         \and     
         Department of Astrophysics, Vietnam National Space Center, Vietnam Academy of Science and Technology, 18 Hoang Quoc Viet, Hanoi, Vietnam
         \and
         Graduate University of Science and Technology, VAST, 18 Hoang Quoc Viet, Cau Giay, Vietnam
         \and
         Institute of Science and Technology Austria (ISTA), Am Campus 1, 3400 Klosterneuburg, Austria
         \and
         Centre for Astrophysics and Supercomputing, Swinburne University of Technology, Hawthorn, VIC 3122, Australia
         \and
         Leibniz Institut für Astrophysik (AIP), An der Sternwarte 16, 14482 Potsdam, Germany
         }

   \date{Received 9th July 2024; accepted 13th July 2024}
 
  \abstract
   {The escape of \lya photons at redshifts greater than two is an ongoing subject of study and an important quantity to further understanding of \lya emitters (LAEs), the transmission of \lya photons through the interstellar medium and intergalactic medium, and the impact these LAEs have on cosmic reionisation.}
   {This study aims to assess the \lya escape fraction, \flya, over the redshift range $2.9<z<6.7$, focusing on  Very Large Telescope/Multi Unit Spectroscopic Explorer (VLT/MUSE) selected, gravitationally lensed, intrinsically faint LAEs. These galaxies are of particular interest as the potential drivers of cosmic reionisation.}
   {We assessed \flya in two ways: through an individual study of 96 LAEs behind the A2744 lensing cluster, with \textit{James Webb} Space Telescope/Near-Infrared Camera (JWST/NIRCam) and HST data, and through a study of the global evolution of \flya using the state-of-the-art luminosity functions for LAEs and the UV-selected `parent' population (dust-corrected). We compared these studies to those in the literature based on brighter samples. }
   {We find a negligible redshift evolution of \flya for our individual galaxies; it is likely that it was washed out by significant intrinsic scatter. We observed a more significant evolution towards higher escape fractions with decreasing UV magnitude and fit this relation. When comparing the two luminosity functions to derive \flya in a global sense, we saw agreement with previous literature when integrating the luminosity functions to a bright limit. However, when integrating using a faint limit equivalent to the observational limits of our samples, we observed enhanced values of \flya, particularly around $z\sim6$, where \flya becomes consistent with 100\% escape. This indicates for the faint regimes we sampled that galaxies towards reionisation tend to allow very large fractions of \lya photons to escape. We interpret this as evidence of a lack of any significant dust in these populations; our sample is likely dominated by young, highly star-forming chemically unevolved galaxies. Finally, we assessed the contribution of the LAE population to reionisation using our latest values for \flya and the LAE luminosity density. The dependence on the escape fraction of Lyman continuum photons is strong, but for values similar to those observed recently in $z\sim3$ LAEs and high-redshift analogues, LAEs could provide all the ionising emissivity necessary for reionisation.   }
  {}
   
   
   \maketitle
%

\section{Introduction}
\label{sect:intro}
For more than 20 years, the study of high-redshift galaxies has been facilitated by the \lya line \citep{cowie1998LAEs, hu1998LAEsearch}. Galaxies for which this line is observable, \lya emitters (LAEs; equivalent width, \ewlya$>25\AA$), have formed an essential part of the drive to characterise star-forming galaxies (SFGs) at increasingly higher redshifts (the most recent examples include, \citealt{Witstok2023lyabubbles,Iani2023lya,Chen2023Lya_escape_JWST}, see also \citealt{Ferrara2023lya_vis} and references therein). The \lya line, while intrinsically very strong and therefore an attractive target for observations, undergoes a very complicated process escaping a galaxy due to the resonant nature of the line \citep{verhamme2008lyadustsim,hayes2010escape_lya, zheng2010lya_radiativetransfer, dijkstra2012lya_transfer, matthee2016calymha_lyaesc} as well as dust within the galaxy \citep{atek2008lya_dust,finkelstein2009dust_emission_lya,hayes2013lyadust}, inflows and outflows \citep{hansen2006lya_transfer,gronke2016LAEoutflows}, and the intergalactic medium (IGM) \citep{stark2010keckLAEfrac,stark2011LAEfrac}. Understanding the escape of these \lya photons is of paramount importance to the characterisation of these high-redshift SFGs as well as to understanding their impact on extragalactic processes such as cosmic reionisation. \\
\indent The escape of \lya photons is also expected to be connected to the escape of ionising Lyman continuum photons ($\mathrm{\lambda_{LyC}<912\,\AA}$) \citep{Dijkstra2016Lya_LyC_UVfaint_reion,verhamme2017lyman,Steidel2018LyC_z3_Lya,Izotov2022LyC_Lya_MgII,Yuan2024LAE_lyC_sim,Choustikov2024Lya_LyC_escape,Pahl2024lya_line_lyC_escape,Gazagnes2024VANDELS_sim_lya_LyC}, although any relation between these two photon escapes tends to be reported with significant scatter between individual galaxies. Since the escape of Lyman continuum photons is extremely challenging to observe directly at redshifts higher than $z\sim3$ due to the intervening IGM, the escape of \lya photons can be an important proxy.\\ 
\indent Many efforts have been made towards a better understanding, both on the theoretical and observational front, and a better quantification of the escape fraction (henceforth \flya). The well-known relations in \cite{kennicutt1998Schmidtlaw} quantify the basic relationship between \lya luminosity and the star-formation rate (SFR) expected to produce it, subject to assumptions on the initial mass function (IMF), stellar metallicity, and star-formation history (SFH). However, the resonance of the \lya line means that \lya photons will scatter in a galaxy's neutral hydrogen. During this process, dust may absorb these photons (\citealt{schaerer2008lya_transfer,ciardullo2014lya_escape,dijkstra2016LAEreview} and references therein), so the \lya escape fraction is expected to decrease with increasing dust content \citep{atek2008lya_dust,Runnholm2020lya_esc_dust}. The size and stellar mass of galaxies may also play a role, with \flya generally increasing with lower stellar mass \citep{oyarzun2017lya_dust,Yang_2017lya_esc_greenpeas,Goovaerts2024}.\\
\indent The increasing neutrality of the IGM at redshifts above $z\sim5-6$ \citep{Mcgreer2018AGN,Planck2018reionisation} is expected to play a significant role in \lya visibility. Neutral hydrogen around the galaxy and along the line of sight absorbs \lya emission, meaning that the number density of LAEs decreases around these redshifts \citep{schaerer2011LAEfrac,pentericci2011laefrac/z=7LBG,deBarros2017LAEfraction,arrabal2018LAEfrac,pentericci2018LAEfrac}. However, this effect is subject to significant uncertainties, and for individual cases, it is highly dependant on the physical conditions and environment of the galaxy in question \citep{goovaerts2023,Witten2023lya_inreionisation}. \\
\indent Despite these challenges, strides have been made in recent years towards quantifying \flya and its evolution with redshift. \cite{blanc2011}, henceforth \citetalias{blanc2011}, studied the evolution of dust properties and \flya across a sample of $\sim100$ blank-field LAEs in the range $1.9<z<3.8$ and compared star-formation rate densities (SFRDs) derived from LAE and UV luminosity functions (henceforth LFs) in order to quantify the evolution of \flya with redshift. They found a sample median of $\sim 29\%$ when considering their LAEs individually and a negligible redshift evolution. \\
\indent \cite{hayes2011redshiftevofdust} found a significantly lower \flya, around $5\%$ at $z=2.2$ using a blind narrow-band survey searching for H$\alpha$ and \lya. The authors made use of the H$\alpha$ line to estimate un-obscured star formation and compared this to the star formation derived from \lya. The H$\alpha$ line originates from the same process as the \lya line and is less sensitive to dust attenuation; therefore, it is often a more reliable indicator of star formation.\\
\indent One general agreement in the literature is that \flya increases with redshift. Low values such as $0-5\%$ are found in the aforementioned study as well as \cite{hayes2013low_z_fesc}, 
 \cite{verhamme2017lyman}, \cite{sobral2017calymha_lya_esc}, \cite{sobral2019predictinglyaesc}, and \cite{Runnholm2020lya_esc_dust}. Larger values such as $30-40\%$ and even higher have been found at redshifts towards the epoch of reionisation (\citetalias{blanc2011};  \citealt{hayes2011redshiftevofdust,Chen2023Lya_escape_JWST,Witstok2024lyaz>8_diff_fesc,Lin2024_lya_escape_Ha_emitters,Napolitano2024lya_visibility_reion}).\\
\indent Another general agreement found in the literature relates to sample selection. Samples that are selected by their \lya emission typically display higher values of \flya when compared to samples selected by H$\alpha$ \citep{song2014hetdex_LAEs_fesc,trainor2016LAEz2.5fesc,sobral2019predictinglyaesc,matthee2021xshooterLAEs}.\\
\indent By contrast, significant scatter in \flya is expected in all samples, likely due to the large number of factors that affect \lya photon escape and the differing conditions present in any sample of galaxies. This can be seen in all studies on \flya, from low redshift \citep{hayes2010escape_lya,hayes2013low_z_fesc,ciardullo2014lya_escape,matthee2016calymha_lyaesc,Runnholm2020lya_esc_dust} to the higher redshift, such as \citetalias{blanc2011}, \cite{Chen2023Lya_escape_JWST}, \cite{Lin2024_lya_escape_Ha_emitters}, and \cite{Napolitano2024lya_visibility_reion}.\\
\indent Currently, the most powerful estimators for determining \flya in an indirect manner include \ewlya \citep{sobral2019predictinglyaesc} and \lya line separation \citep{Yang_2017lya_esc_greenpeas,matthee2021xshooterLAEs}. The correlation with \lya line separation is likely due to the presence of outflows, to which the \lya line shape is sensitive \citep{gronke2016LAEoutflows,Blaizot2023lya_line}. These outflows produce channels of low HI column density in the interstellar medium (ISM) and circumgalactic medium (CGM), which allows for more \lya escape \citep{Hashimoto2015lya_profile_HIdensity}. \\
\indent We can also constrain \flya in a global way by comparing the SFRDs derived from the LAE and general UV-selected populations (with SFR derived from the UV emission). The ratio of these gives the escape fraction of \lya photons over a specific survey volume. Employing this method, \cite{Konno_2016}, \citetalias{blanc2011} and \cite{hayes2011redshiftevofdust} found a gentle evolution over their redshift ranges that could be fitted with a power law of the form: \flya = $C\times (1+z)^{\xi}$, where $C=(4.79^{+5.68}_{-0.69}) \times 10^{-4}$, $\xi=3.38^{+0.10}_{-0.37}$ at $z\sim 2.2-6.7$ \citep{hayes2011redshiftevofdust}, $C= 3.98^{+2.32}_{-1.46} \times 10^{-3}$, $\xi=2.2\pm0.3$ at $z=1.9-3.8$ \citepalias{blanc2011} and $C=5.0 \times 10^{-4}$, $\xi=2.8$ in $z=0-6$ \citep{Konno_2016}.\\
\indent \cite{hayes2011redshiftevofdust} mentioned that \flya may reach unity at $z\sim 11 ^{+0.8}_{-0.6}$, while \cite{cassata2011vimosLAELF} argued that \flya would reach unity around a redshift of six. In general, such research has been done with a few LAEs fainter than $\mathrm{log\, L_{Ly\alpha}} \, \text{[erg/s]} < 40 $, hence without the ability to fully constrain the faint part of the LAE LF and its impact on the SFRD.\\
\indent Any difference in the evolution of \flya may come from the integration limits during the estimation of the \lya and UV luminosity densities. Hence, data on the fainter luminosity regime is necessary in order to integrate to fainter luminosity limits and better understand these faint populations' effect on the evolution of the LF as well as \flya with redshift.\\
\indent Fortunately, the recent VLT/Multi Unit Spectroscopic Explorer (MUSE) observations of lensing clusters \citep{AC2022LLAMAS,Thai2023} and blank fields \citep{Vitte2024} have allowed us to observe  \lya down to $\mathrm{log\,L_{Ly\alpha}} \, \text{[erg/s]} \sim 39 \text{,}\, 39.5$, respectively.    \\
\indent In this paper, we seek to combine two state-of-the-art LFs of intrinsically faint lensed galaxies, the latest UV LF from \cite{Bouwens2022UVLF2z9} and the latest LAE LF from \cite{Thai2023}, allowing us to compare the SFRD derived from \lya emission, $\mathrm{SFRD_{Ly\alpha}}$, to the total SFR density, $\mathrm{SFRD_{total}}$, and hence track the redshift evolution of \flya over a wide redshift range: $2.9<z<6.7$. In \cite{hayes2011redshiftevofdust}, the independent LF comparison is advocated as a method to calculate \flya, though it is subject to cosmic-variance related errors. We can significantly reduce the uncertainties related to cosmic variance by using six lensing clusters for the UV LF and 17 lensing clusters for the LAE LF. The LFs we used provide this study with a unique opportunity to explore the faint galaxy regime ($\mathrm{M_{UV}<-13}$ and $\mathrm{log\,L_{Ly\alpha}\,[erg/s]>39}$) without the need for extrapolation in the LFs.\\
\indent We also calculated \flya on an object-per-object basis in order to compare galaxy properties as well as the global \flya evolution, based on spectral energy distribution (SED) fitting of HST and JWST/NIRCam data of $\sim100$ LAEs in the A2744 lensing cluster. The NIRCam data, extending to $\sim5\,\mu m$, allowed us to access, for the first time, the rest-frame optical emission of these LAEs and greatly improves the reliability of the properties extracted with SED fitting. The NIRCam imaging notably also covers the entire MUSE field of view, meaning the full sample size of MUSE LAEs can be explored. This would not be the case with calculations of SFR from H$\alpha$ and dust attenuation from the Balmer decrement obtained from spectroscopy, for example from JWST/NIRSpec's MSA mode. This would, in principle, be preferable, due to the fewer assumptions needed and our lack of photometric coverage of the infrared for our LAEs. However, samples of LAEs with Balmer line detections are rare at the redshifts concerned by this study, and sample sizes are often small. Using photometry, we could gain an appreciable sample size using just one cluster, although more would be preferable (and soon possible with the current and upcoming JWST observations). \\
\indent Studies of emission lines with slit spectroscopy for these galaxies (especially lensed LAEs and therefore potentially extended sources, both by lensing and due to the extended nature of \lya haloes) are not without difficulties themselves. Parts of flux can be missed in slits, especially slits as small as JWST/NIRSpec's MSA ($0\farcs2$ to $0\farcs45$; \cite{Napolitano2024lya_visibility_reion} provide a good discussion of these difficulties, see also \cite{Jung2024_LAEs_missed_slit}), and sometimes, flux can be missed altogether, such as in \cite{Jiang2023_lya_nondetection}.\\ 
\indent Section~\ref{sect:individual_escape} outlines the determination of \flya for individual galaxies, describing the data and process used. In Section~\ref{sect:global_escape}, we describe the global evolution of \flya , the two LFs used for this calculation, and the results thereof. A comparison of the two methods to each other and to studies in the literature is presented in Section~\ref{sect:Discussion} along with consequences for reionisation and a discussion of the validity of comparing the UV and LAE LFs. Finally, Section~\ref{sect:conclusion} offers conclusions and an outlook for \flya determination in the near future.\\
\indent Throughout this paper, we adopt a value for the \textit{Hubble} constant of $H_0=70\,\mathrm{km\,s^{-1}Mpc^{-1}}$, and the cosmology used is $\Omega_{\Lambda}=0.7$ and  $\Omega_{\mathrm{m}}=0.3$. The adopted IMF is that of \cite{salpeter1955IMF}. All values of \lya luminosity and absolute UV magnitude are given corrected for magnification.

\section{Escape fraction of individual LAEs}
\label{sect:individual_escape}

\subsection{Data: Combining spectroscopy and photometry}
\label{sect:data}
In order to perform analysis of this genre, photometry and spectroscopy of sufficient quality and depth is essential. For our purposes we combined public MUSE integral field unit (IFU) spectroscopic data \citep{richard2021atlas} (094.A-0115, 095.A-0181, 096.A-0496) with the latest, deepest combination of \textit{Hubble} Space Telescope (HST) and \textit{James Webb} Space Telescope (JWST) photometry from the UNCOVER survey (PIs Labbé and
Bezanson, JWST-GO-2561, \cite{Bezanson2022UNCOVER}) of the lensing cluster A2744. This cluster is extremely well studied and benefits from a great number of multiple images to constrain its lens models. Below, we detail the two sets of observations and how they are used for this study.

\subsubsection{MUSE IFU spectroscopy}
\label{sect:MUSE}
The use of an IFU allowed us to blindly select LAEs, rather than performing spectroscopic follow-up on UV-selected targets. This ensures our sample has a simple  reproducible selection function, namely \lya luminosity limited. MUSE has a field of view of $1\times1\,\mathrm{arcmin}^2$ and a spectral resolving power of $R \sim 3000$ at $\lambda=808\,\mathrm{nm}$ \citep{bacon2015muse_details}. Wavelength coverage ranges between $4750\,\AA$ and $9350\,\AA$, meaning the \lya line can be detected between redshifts of 2.9 and 6.7.\\
\indent The A2744 lensing cluster used in this work has integration times varying between 3.5 and 7 hours. It forms part of a data release by \cite{richard2021atlas}\footnote{\url{https://cral-perso.univ-lyon1.fr/labo/perso/johan.richard/MUSE_data_release/}} and the LAEs used in this work are comprehensively described as part of the Lensed \lya MUSE Arcs (LLAMA) Sample, in \cite{AC2022LLAMAS}.The full process of LAE detection in MUSE data is detailed in \cite{richard2021atlas} and \cite{PW2020MUSEdata}. We give an abridged version here. \\
\indent Emission line sources are identified in MUSE narrow-band datacubes using the MUSELET software \citep{piqueras2019asp}\footnote{\url{https://mpdaf.readthedocs.io/en/latest/muselet.html}}. Thereafter the Source Inspection package \citep{bacon2022musedatareleaseII} is used to identify line emitters as LAEs and assign redshifts. During this process, the redshift confidence of each LAE is determined, based on the emission lines seen, the shape of the \lya line, ancillary HST data and lensing considerations. The confidence scale ranges from 1 for a tentative redshift attribution to 3 for a very secure redshift attribution. In this work, only sources that have been identified with redshift confidence levels of two and three are used, as well as a S/N greater than three. \\
\indent The flux of these LAEs is derived using SExtractor \citep{EB96SEx} on a continuum-subtracted Narrow-Band sub-cube with a size of $15"\times 15"\times 20\AA$. The full process is described in \cite{GdlV2019LAELF}.

\subsubsection{HST + JWST photometry and galaxy properties}
\label{sect:photometry}
The photometric catalogues of the UNCOVER project \citep{Weaver2023UNCOVER_catalogs} comprise photometry in 15 bands ranging from the F435W HST filter to the F444W JWST/NIRCam filter. This gives excellent, deep coverage up to 4.4\mm, which corresponds to $\sim580\,\mathrm{nm}$ in the rest frame at $z=6.7$, the highest redshift of the MUSE LAEs. We used the catalogue optimised for faint and compact sources ($0.32\arcsec$ apertures). $5\sigma$ depths in the 15 filters range between 27.16 and 29.56 in AB
magnitudes. Of the 154 total LAE images, 121 were detected, which corresponds to 99 individual LAEs. We rejected two of these LAEs as being clearly contaminated by bright cluster galaxies and a further LAE, as it lies on a critical line and has a magnification factor of $137\pm1500$. As such, its properties would likely not be well constrained (see, for example, \citealt{Limousin2017_critical_lines}). Having thus blindly matched these \lya-selected galaxies to their UV counterparts, their SEDs were fitted with the SED-fitting code {\tt CIGALE} \citep{boquien2019cigale}. The details of the fitting procedure are outlined in \cite{Goovaerts2024}; here we give a brief outline.\\
\indent Taking advantage of the secure spectroscopic redshifts of these LAEs, we fit their SEDs using two different SFHs: a single exponentially decaying burst with a range of timescales and a double burst model with a second, delayed burst at a time $\tau$ after the initial burst, to be fit \citep{malek2018cigaleSFH}. Ages for the bursts range from 10 to 700$\,\mathrm{Myr}$ and mass fractions in the second burst range from 0.001 to 0.65. The stellar template library of \cite{BandC03SED} is employed with metallicities ranging from 0.001 to 0.02, together with the nebular emission models from \cite{Theule2024SEDfitting_photodiss}. Gas metallicities range between 0.0004 and 0.02. Dust attenuation is taken into account by means of the Calzetti attenuation law \citep{Calzetti00dust}. \\
\indent The best-fit SFH and model was carried forward in this analysis (see \citealt{Goovaerts2024} for comparisons between the two SFHs). One of the parameters in the {\tt CIGALE} output is the intrinsic dust-corrected star formation, which is used in the computation of \flya. 

\subsubsection{Lensing models}
\label{sect:lensing_models}
For studies involving lensing clusters, the lens models used are pivotal to the reliability of the results, as these govern the lensing magnifications that are assigned to each galaxy and therefore the correction made to their observed luminosity. The lens models used for this work for the A2744 lensing cluster are described in \cite{richard2021atlas}. As the Hubble Frontier Fields (HFF) and A2744 in particular have been so well studied over the years, these clusters have very well constrained mass models, drawing on many multiple images identified behind each cluster. While systematic uncertainties are still associated to the particular lens model used \citep{Acebron2017lensing_systematics,Furtak2021cluster_stellasmassfunc}, the quantity of multiple images for these particular clusters render the statistical error on the mass distribution as low as $1\%$ \citep{richard2021atlas}. Typical, accepted magnifications in the sample range from 1.5 to 25, with the maximum magnification being 137. We note that escape fraction measurements for individual galaxies (as in the following Section), are not affected by magnification. 

\subsection{Escape fraction}
\label{sect:obj_esc}
\begin{figure}
    \centering
    \includegraphics[width=0.5\textwidth]{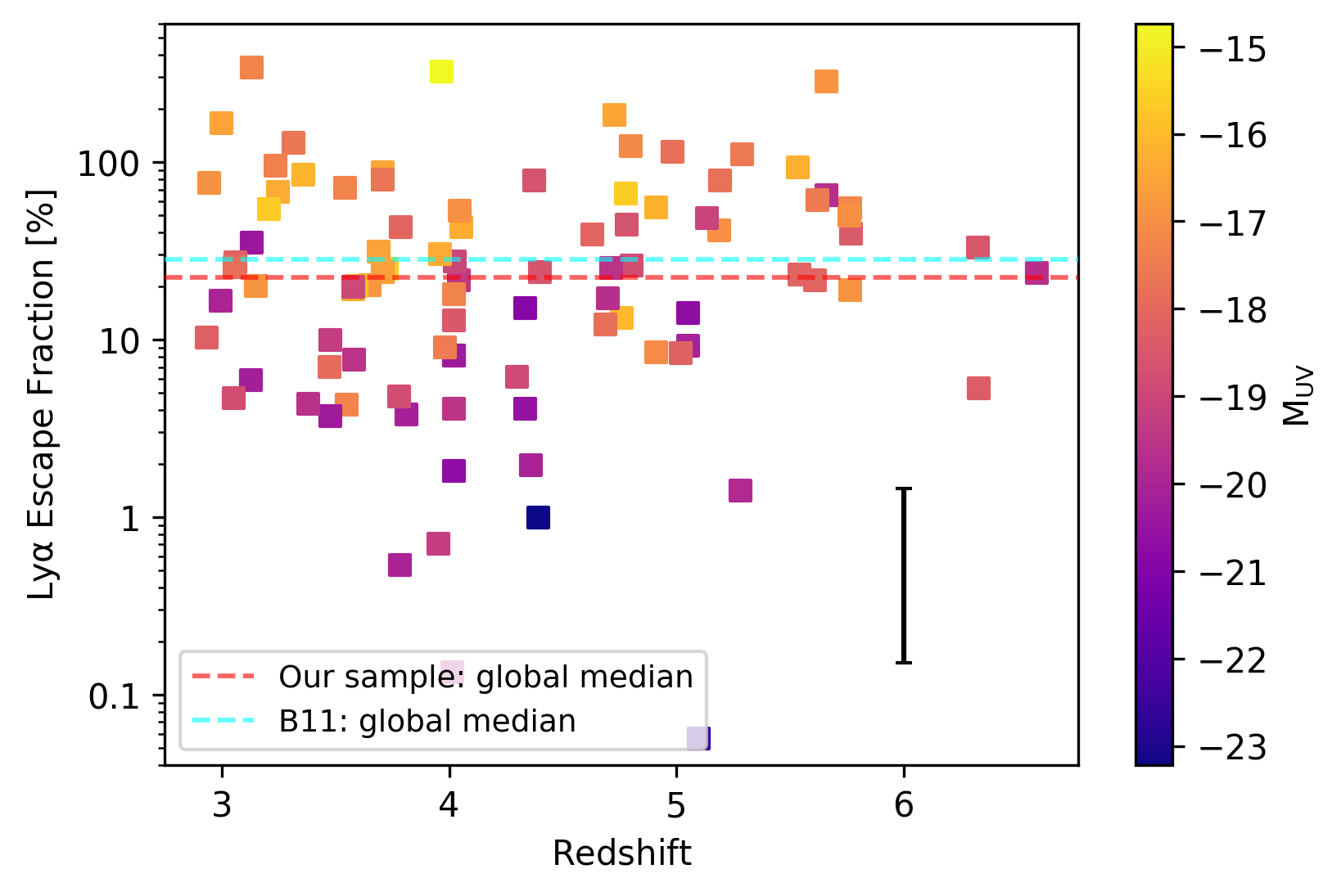}
    \caption{Redshift evolution of the escape fraction of \lya photons from LAEs in the A2744 lensing cluster colour-coded by the absolute UV magnitude of each LAE. The global medians of our sample and that in \citetalias{blanc2011}, are shown with red and cyan dashed lines respectively. The median uncertainty of the sample ($\sim18\%$) is shown by the black error bar, as if it was at the level of the global median, offset for clarity.}
    \label{fig:Fesc_individual}
\end{figure}

\begin{figure}
    \centering
    \includegraphics[width=0.5\textwidth]{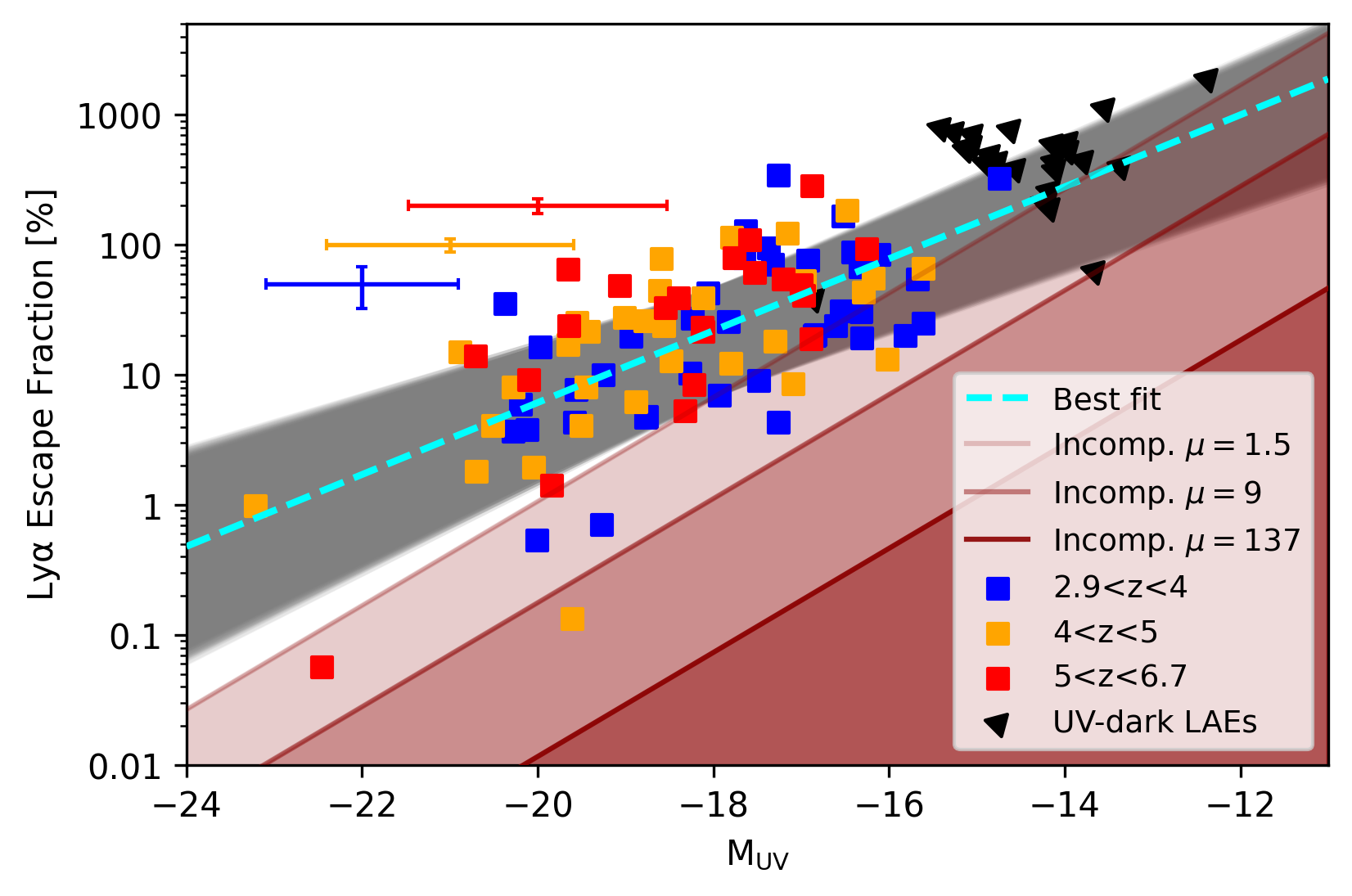}
    \caption{Evolution of \flya with absolute UV magnitude $\mathrm{M_{UV}}$. Photometrically detected galaxies are split into three redshift bins to stress the lack of evolution in this sense, while the evolution with $\mathrm{M_{UV}}$ is fit by the relation in cyan (details in text). Median error bars for the three redshift bins are shown in the corresponding colour and \flya error bars are also listed here for clarity (relevant redshift bin in brackets): $0.18 \,(2.9<z<4)$, $0.12 \,(4<z<5)$ and $0.25 \,(5<z<6.7)$. Absolute magnitude values and their errors are adjusted for magnification. Black triangles denote continuum-undetected LAEs. $\mathrm{M_{UV}}$ values are upper limits and escape fraction values are lower limits. These objects are not included in the fit. Red shaded areas denote regions of incompleteness based on MUSE line-detections: a darker colour representing increasing lensing magnification (see text).}
    \label{fig:MUV_fesc}
\end{figure}

To calculate the escape fraction on an object-per-object basis for the 96 individual LAEs behind the A2744 lensing cluster, from the data described in Section~\ref{sect:data}, we compare the SFR inferred from the \lya flux, to the dust-corrected SFR from the {\tt CIGALE} SED fitting process. The SFR inferred from the \lya flux is calculated using the prescription in \cite{kennicutt1998Schmidtlaw} and the conversion factor of 8.7 between \lya and H$\alpha$ luminosities, using a Case B recombination scenario \citep{Osterbrock_1989} ($T=10^4\,\mathrm{K}$). \\
\indent The ratio of these two SFRs gives the escape fraction of \lya photons, \flya. The redshift evolution of this quantity is shown in Fig.~\ref{fig:Fesc_individual}. The sample median is shown as the red dashed line and the sample median in \citetalias{blanc2011} is shown as the cyan dashed line. They are consistent to within each other's errors, even though \citetalias{blanc2011} use a different prescription for finding \flya. The median escape fraction uncertainty of our sample ($18\%$) is shown by the black error bar. Each LAE is colour-coded by its absolute UV magnitude, calculated from the filter that sees the rest frame $1500\,\AA$ emission. One can clearly see that \lya photons escape more easily from UV-fainter objects, a result already remarked upon by \cite{Lin2024_lya_escape_Ha_emitters}, \cite{Napolitano2024lya_visibility_reion}. To clarify this further, we can fit a relation between the absolute UV magnitude and \flya for each galaxy. This is shown in Fig.~\ref{fig:MUV_fesc}, where the galaxies are now colour-coded by redshift-bin and median error bars for each bin are shown in the corresponding colour. The best-fit to this data is
\begin{equation}
    \mathrm{log(\flya)=(0.27\pm0.02)\,M_{UV}+(4.2\pm0.5)}.
\end{equation}
\indent This is clearly a tighter relation than the one shown by the redshift versus \flya, with an uncertainty on the slope and normalisation of about $10\,\%$. These quantities, $\mathrm{M_{UV}}$ and \flya are not independent as $\mathrm{M_{UV}}$ is related to the intrinsic SFR of a galaxy, so this relation is to be expected, but it is nevertheless noticeable that below $\mathrm{M_{UV}}=-18$ there are few galaxies with \flya values less than 10\%. However, care must be taken when interpreting this result, as incompleteness can play a role as we start to probe fainter galaxies (towards the right of Fig.~\ref{fig:MUV_fesc}). To better assess this, in Fig.~\ref{fig:MUV_fesc} we plot three regions where we expect to be incomplete as a function of UV magnitude and lensing magnification. This is based on the limiting flux for a MUSE detection (of an LAE). The three magnification regions plotted are $\mu=1.5$, the minimum magnification included in the sample (the area covered by MUSE observations do not reach the areas with less magnification than this, most of the area has $\mu>2$), $\mu=9$, the maximum magnification for an image included in the sample (we note that there are LAEs not included and images not included with $\mu>9$) and $\mu=137$, the maximum magnification of any LAE image in the A2744 lensing cluster. These three regions delineate the area of this plot where we expect to become incomplete in terms of detecting a faint galaxy with a small \flya value with MUSE. We can see that this effect likely becomes important towards the faint end of this plot, at $\mathrm{M_{UV}\gtrsim-17}$. However, at the brighter end of this graph, we do not expect the incompleteness to shape the relation we observe.\\ 
\indent We did see LAEs undetected in the continuum that might be expected to fill this region of the graph. However, they are uniquely objects of high escape fraction (based on upper limits of the UV continuum, and no dust correction). These objects are denoted by the black triangles in Fig.~\ref{fig:MUV_fesc}: lower limits on \flya and upper limits on $\mathrm{M_{UV}}$.\\
\indent We note also that around this $\mathrm{M_{UV}}$ value of $-18$ that we start to see objects with \flya$>100\%$. \flya$>100\%$ means that we are seeing more \lya emission than we should, given the standard conversions between \lya and UV luminosities and SFR detailed above. Reasons for this may include our assumptions in making those conversions, such as constant SFH, ages greater than $\sim 50\,$Myr and the case B recombination scenario \citep{Osterbrock_1989}, becoming less valid (see, for example, \citealt{McClymont2024caseB?}). This may point concurrently to our models failing to correctly estimate the dust content of these galaxies and thereby underestimating the intrinsic SFR. However, this would mean that some property of these galaxies is heavily attenuating the UV emission, but not attenuating the \lya emission accordingly. \cite{atek2008lya_dust} postulate that geometry or the presence of large outflows may play a role in disconnecting \lya emission and dust such that one may see a high \flya from a heavily dust-attenuated system.  \\
\indent We also observed LAEs for which the continuum is completely undetected in the deepest photometry to date \citep{bacon2015muse_details,Maseda2020faintLAEs,GdlV2020LAEfrac,goovaerts2023,Maseda2023UVdarkLAEs}, and \flya for these galaxies is, naturally, very high, often greater than 100\% (as calculated from $2\sigma$ upper limits on the UV continuum). This suggests that there is more than simply model inadequacy at play here. We discuss these objects further in Section~\ref{sect:Discussion}.  \\
\indent Unlike $\mathrm{M_{UV}}$, we see no significant evolution of \flya with redshift, although we have relatively few objects at redshifts higher than 6. It is clear from Fig.~\ref{fig:Fesc_individual} that we are affected by missed objects above $z=6$, both in the high-\flya regime (as these objects have low $\mathrm{M_{UV}}$) and the low-escape fraction regime, as objects with low \flya are harder to detect at higher redshifts. The greater presence of skylines and decreased MUSE sensitivity towards longer wavelengths are additional challenges to detect LAEs towards the high end of our redshift range. \\
\indent The comparison to the results in \citetalias{blanc2011} suggest that \flya does not evolve significantly over the redshift range $1.8<z<6$. This is in tension with certain literature results such as \cite{hayes2011redshiftevofdust}, who find that \flya increases with redshift ($0.6<z<6$). However, this study was performed in a similar way to the global analysis of \flya that we present in the following sections. Hence, we return to this point after having taken both methods into consideration.\\
\indent The scatter in the individual results shown in Fig.~\ref{fig:Fesc_individual} is to be expected, given the wide variety of conditions within galaxies, even LAEs, expected to be mostly young and highly star forming. Dust content of these galaxies likely varies, which plays a significant role in the escape of \lya photons \citep{verhamme2008lyadustsim}. Additionally and as mentioned, geometry and line-of-sight effects are also expected to have a hand in the amount of \lya photons that reach the observer \citep{Verhamme_2012,dijkstra2012lya_transfer,gronke2016LAEoutflows,Blaizot2023lya_line,Giovinazzo2024lya_reion_sim}.\\
\indent While we see consistency between the sample median of \citetalias{blanc2011} and ours, the offset, as well as the smaller scatter of our results, is to be expected based on the differing methods for finding \flya. \citetalias{blanc2011} assume an intrinsic UV slope of $-2.23$ in order to calculate the dust reddening $E(B-V)$. As this UV slope likely depends on factors other than dust, such as metallicity, age and mass of the stellar population \citep{buat2012dust_UVslope,wilkins2013UVslope,bouwens2016UVslope_stellarmass_IRexcess,reddy2018UVslope_dust,Topping2022blue_UVslopes}, and as there is evidence for slopes bluer than this value \citep{deBarros2017LAEfraction,Naidu2022luminous_early_gals,goovaerts2023,Iani2023lya,Nanayakkara2023blue_UVslopes,Napolitano2024lya_visibility_reion}, we prefer not to use this assumption. If there are bluer intrinsic slopes than this, we would expect the corresponding $E(B-V)$ values from \citetalias{blanc2011} to be higher, and therefore, the intrinsic SFR would be higher, and the calculated escape fractions would be lower. This is what we observed in our results. We also saw less scatter in \flya, as we do not depend on estimations of the UV slope, which are known to have significant scatter (as well as uncertainty and possibly bias, e.g. \citealt{Austin2024UVslopes}).\\
\indent Our sample median, $22.5\,\%$, agrees with determinations at slightly lower redshift from \cite{sobral2017calymha_lya_esc,matthee2021xshooterLAEs}, who both quote escape fractions of $\sim30\,\%$ for LAEs. Most recently, also using JWST, \cite{Chen2023Lya_escape_JWST} have found a sample median of $28\,\%$ for 10 LAEs at $5<z<8$.\\
\indent However, we note that the scatter on individual determinations of \flya remains a very significant consideration when assessing samples of the kind in Fig.~\ref{fig:Fesc_individual}. It is therefore unlikely that we will be able to specify a precise fraction for \lya escape at any given redshift. It is clear that the trend of \flya with $\mathrm{M_{UV}}$ is more useful. \\
\indent We note also that while individual \flya values are better determined using H$\alpha$ fluxes and the Balmer decrement, the intrinsic scatter seen remains similar, with values ranging from $<1\%$ escape to $100\%$ (see, for example, \citealt{Chen2023Lya_escape_JWST,Napolitano2024lya_visibility_reion}). For our sample, we can take additional encouragement from the fact that the dust content of these low-mass LAEs, at these redshifts, is both expected to be, and observed to be, very low. We therefore do not expect many cases where our SED fitting significantly underestimates the dust attenuation of a galaxy. Both methods of determining \flya are valuable then, and a more detailed comparison between them for the same sample of galaxies is desirable.

\section{Evolution of the global \lya escape fraction}
\label{sect:global_escape}

\begin{figure*}
    \centering
    \includegraphics[width=0.49\textwidth]{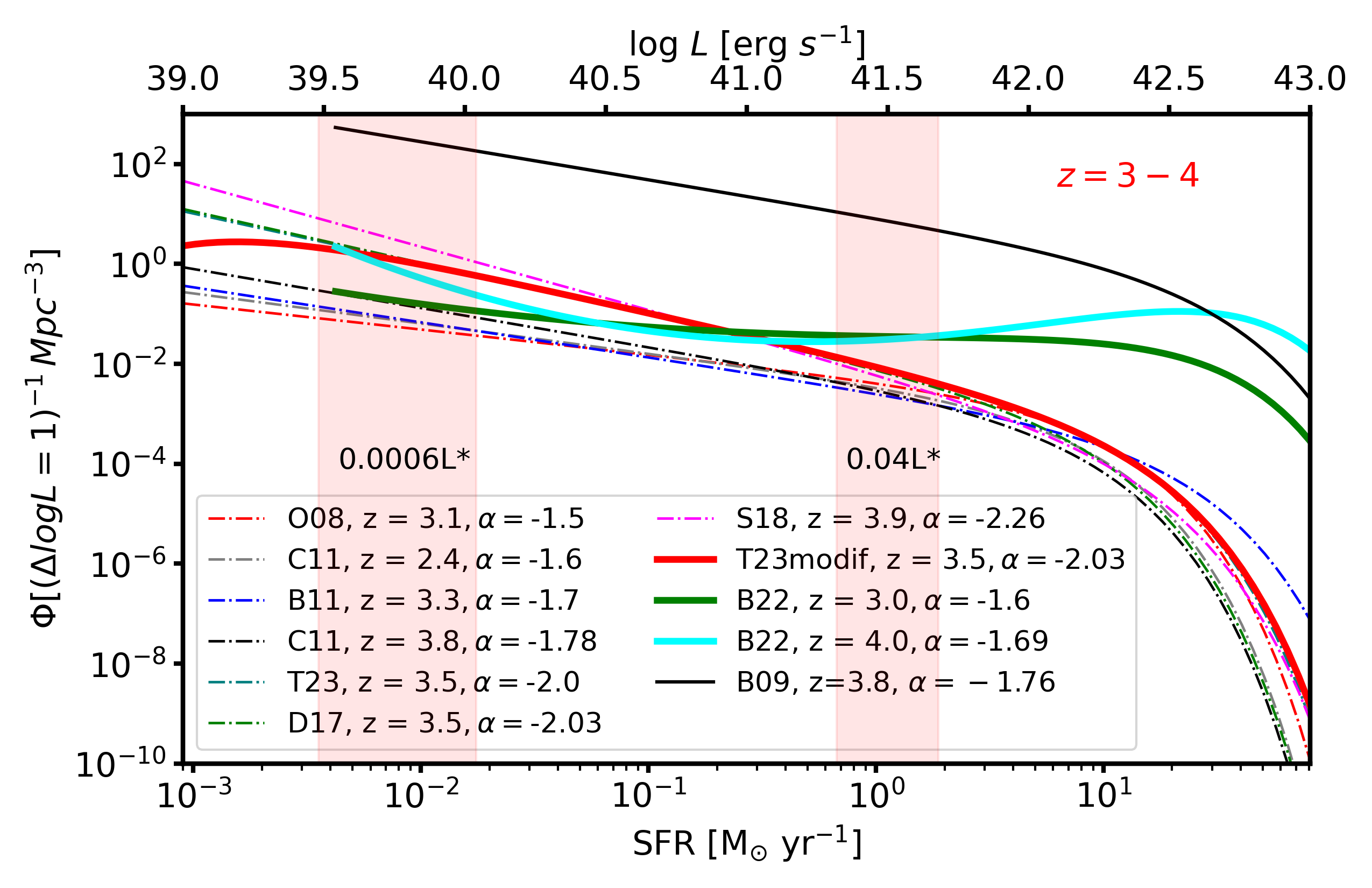}
    \includegraphics[width=0.49\textwidth]{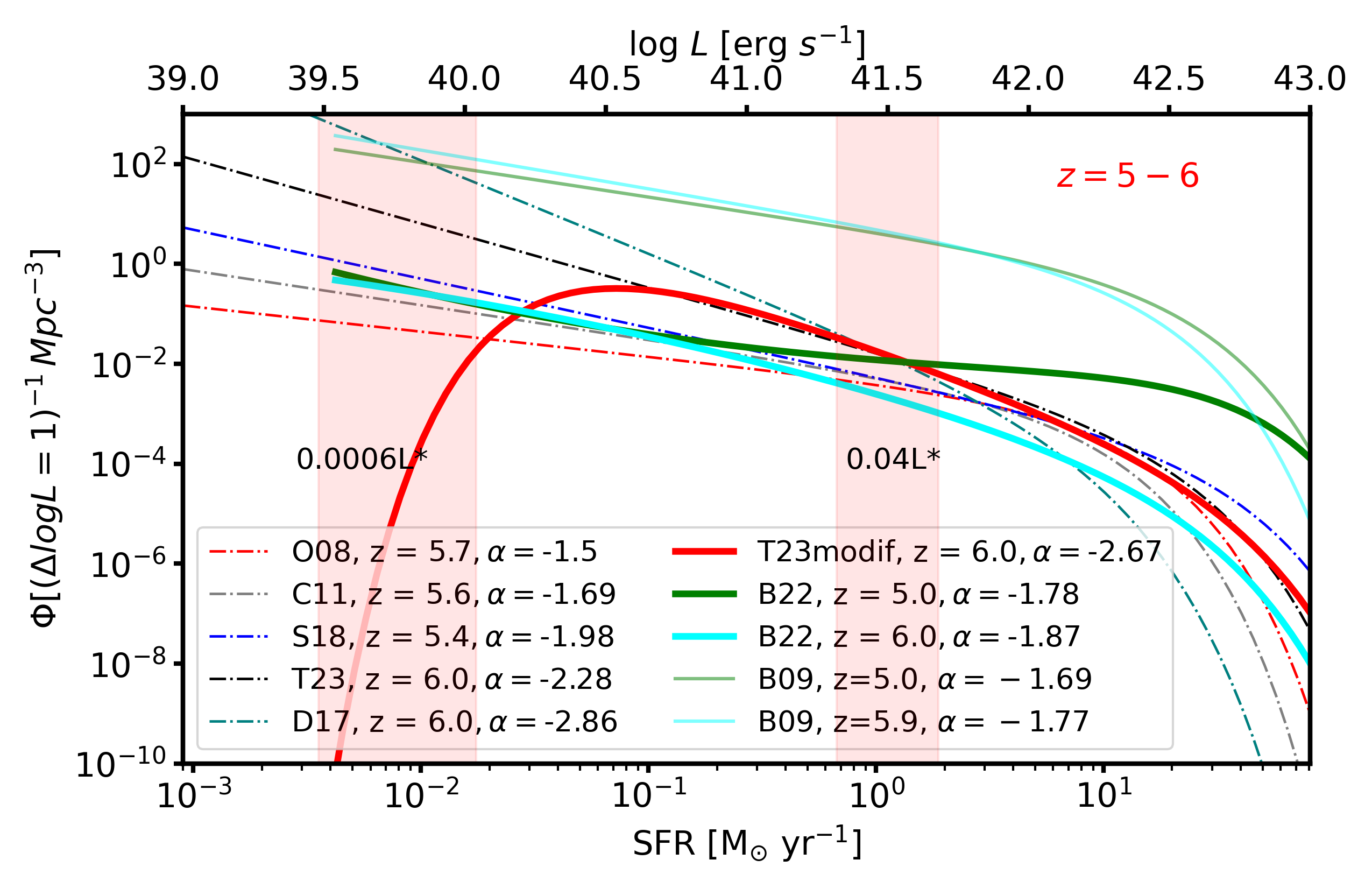}
    \caption{Ultraviolet and LAE LFs included in our analysis on a common x-axis of SFR. The LFs of \cite{Bouwens2022UVLF2z9} and \cite{Thai2023} are shown in bold and the literature LFs are shown in coloured dash-dotted lines. The LF integration limits in terms of $L*$ that we use throughout this work are denoted by the pink shaded regions.  \textit{Left:} Results from the redshift range 3 to 4. \textit{Right:} Results from the redshift range 5 to 6. The literature LFs in the graph are shown in shorthand in the caption and here in full, in order: \cite{Ouchi2008LAE_LF,cassata2011vimosLAELF,Sobral_2018,Thai2023,drake2017MUSELAELF,Bouwens2022UVLF2z9,Bouwens2009UV_LF_SFRD}.}
    \label{fig:all_LFs}
\end{figure*}

One can derive estimates of \flya in a global manner by comparing the LF of LAEs and that of the `total' galaxy population. Integrating both of these LFs gives the \lya luminosity density \rlya and the total luminosity density \rtot. From this one can derive the respective SFRDs. Then taking the ratio of these SFRDs gives \flya \citep{hayes2011redshiftevofdust, blanc2011, Zheng_2013, Konno_2016}. In order to perform this calculation, we start from the two state-of-the-art LFs in lensing fields, the LAE LF from \cite{Thai2023} and the UV LF from \cite{Bouwens2022UVLF2z9}, assumed here to be the `total' galaxy population. This assumption and its consequences for the results of this study are discussed in detail in Sect.~\ref{sect:undetected_LAEs}.\\
\indent Before presenting the derivation of \flya, we briefly detail both LFs used, starting with the LAE LF from \cite{Thai2023}.

\subsection{\lya luminosity function}
\label{sect:LAE_LF}
The LAE LF is computed by using the non-parametric $V_{\text{max}}$ method to estimate the volume of the survey in which we can detect sources individually. The inverse of this parameter is the contribution of the source to the numerical density of the galaxies. The procedure to compute this parameter using data collected by lensing observations by MUSE/VLT was developed by \cite{GdlV2019LAELF} and presented in \cite{Thai2023} with some improvements related to the magnification and completeness values of the sources. These completeness values reflect the expected number of galaxies actually in the field based on the observed galaxies and the noise statistics of the field. The sample in \cite{Thai2023} includes 600 LAEs behind 17 lensing clusters observed by MUSE/VLT, identified with highly secure redshifts, $z_{conf} = 2,3$ (high \lya line S/N, clearly asymmetric line profile, etc) as defined in \cite{richard2021atlas}. These data are within a redshift interval $2.9<z<6.7$ and covering four orders of magnitude in luminosity $39< \text{log} \, L\,[\text{erg} \text{ s}^{-1}]<43$. These luminosity values were corrected for the magnification of the lensing effect using models from the work of \cite{pello1991, covone2006, richard2010, mahler2018, lagattuta2019, beauchesne2023AS1063lensmodel, richard2021atlas}. \\
\indent The evolution of the LF with redshift was investigated in four redshift intervals $2.9<z<6.7$, $2.9<z<4.0$, $4.0<z<5.0$ and $5.0<z<6.7$ (labelled $z_{all}$, $z_{35}$, $z_{45}$ and $z_{56}$) and the Schechter function used to fit the LF points. Using the data sample collected with the help of gravitational lensing is efficient to study the LF at faint \lya luminosities (log $L$ [erg s$^{-1}$] < 41). However, it becomes less sensitive when examining the function surrounding the characteristic luminosity value $L*$ of the Schechter function. To tackle this problem, \cite{Thai2023} used the average LF values obtained from the literature in the same luminosity and redshift bins as a constraint at the bright end of the LF. A turnover in the shape of the LF was observed at luminosities log $L$ [erg s$^{-1}$] < 41 when studying the evolution in the $z_{45}$ and $z_{56}$ redshift intervals. This was explained by inefficient cooling of gas in small dark matter haloes \citep{jaacks2013_turnover, gnedin2016_turnover, yue2016_turnover}. To accommodate the potential turnover in these two redshift bins, a modified Schechter function was introduced by multiplying an exponential term $\text{exp} -(L_T/L)^m$ to the traditional Schechter function. The best-fit results suggest that $m$ is about unity and log $L_T$ [erg s$^{-1}$] is $\sim 40$ and $\sim 40.7$ for the redshift intervals $z_{45}$ and $z_{56}$ respectively.  Accounting for the effects of source selection, completeness cut, and flux measurement, the final slope values in the four redshift intervals are $-2.00\pm0.50$, $-1.97\pm0.50$, $-2.28\pm0.50$ and $-2.06\pm0.60$. These values are in good agreement with results from other literature studies within a $1\sigma$ deviation.  The best-fit value of the characteristic luminosity values are log $L*$ [erg s$^{-1}$] = $42.85^{+0.10}_{-0.10}$, $42.87^{+0.11}_{-0.1}$, $42.97^{+0.13}_{-0.11}$, $43.09^{+0.10}_{-0.08}$ and the normalisation factors are $\phi*[10^{-4}\text{Mpc}^{-3}]$ = $7.41^{+2.70}_{2.70}$, $6.56^{+3.20}_{-2.40}$, $4.06^{+2.70}_{-1.70}$ and $3.49^{+2.11}_{+1.50}$ in redshift bins $z_{all}$, $z_{35}$, $z_{45}$ and $z_{60}$, respectively.  \\
\indent In our analysis we also include results for the LAE LF at different redshift intervals from literature studies to build a general picture, and provide brief descriptions of these works here.  \cite{GdlV2019LAELF} studied the LF evolution of $\sim$ 120 LAEs behind four lensing clusters observed by MUSE, with the same redshift range as \cite{Thai2023} and a luminosity range of $39\lesssim\mathrm{log\,L_{Ly\alpha}[erg/s]\lesssim43}$. In the blank field observations, \cite{Ouchi2008LAE_LF, blanc2011, cassata2011vimosLAELF, drake2017MUSELAELF, Konno_2016} presented the LF evolution with redshift of LAEs with luminosities greater than log $L \,\text{[erg/s]} = 41$. Due to lacking data at faint luminosities, \cite{Ouchi2008LAE_LF} fixed the slope value of the Schechter function at $\alpha=-1.5$ ($z=3.1, 3.7, 5.7$). \citetalias{blanc2011} fixed this parameter at $\alpha = -1.7$ ($z=1.9-3.8$), while \cite{cassata2011vimosLAELF} found $\alpha=-1.6^{+0.12}_{-0.12}$, $-1.78^{+0.10}_{-0.12}$, $-1.69$ (fixed) at $z=1.95-3.0, 3.0-4.55, 4.55-6.6$, respectively. However, at a higher redshift and a brighter luminosity range, this value has been found to be much steeper, varying from $-2.26$ to $-2.56$ at redshift $z=5.7$, and from $-1.86$ to $-2.5$ at redshift $z=6.6$ \citep{Konno_2016}; or $\sim -1.98$ at redshift $z=5.4\pm0.4$ \citep{Sobral_2018}, depending on the luminosity limitations used for fitting.

\subsection{Ultraviolet luminosity function}
\label{sect:UV_LF}

\cite{Bouwens2022UVLF2z9} presented a new derivation of the UV LF in the redshift range $2.0< z< 9.0$ including over 2500 lensed galaxies behind the HFF clusters using HST observations from the HFF \citep{lotz2017frontier}, Grism Lens-Amplified Survey from Space (GLASS) \citep{treu2015GLASS} and Cluster Lensing And Supernova survey with \textit{Hubble} (CLASH) \citep{Postman2012CLASH} programmes. This LF probes down to $\mathrm{M_{UV}}<-16$ over the full redshift range we are considering in this study. A smooth flattening of $\alpha$ was found, from $\alpha=-2.28\pm0.10$ at redshift $z=9$ to $\alpha=-1.53\pm0.03$ at redshift $z\approx2$. Over our redshift range, $3\lesssim z\lesssim7$, the faint-end slope is found to evolve from $\alpha\sim-1.60$ to $\alpha\sim-2.05$. The shape of the LF and its evolution agrees well with other results in the literature, including determinations from blank fields (see \citealt{Parsa2016UVLF}, as well as other references in \citealt{Bouwens2022UVLF2z9}).
\\
\indent We treat this LF as the LF of the total ``parent'' population in order to calculate the total SFRD. In this way, inclusive of completeness corrections, we assume that both LFs represent the ``total'' of their respective populations within the quoted \lya luminosity and absolute UV magnitude limits. This is an assumption that we return to and discuss in Section~\ref{sect:undetected_LAEs}.\\
\indent All LFs, UV and LAE, are shown on a common scale (SFR) in Fig.~\ref{fig:all_LFs}. The LFs of \cite{Thai2023} and \cite{Bouwens2022UVLF2z9} are highlighted in thicker, solid lines.

\subsection{Global evolution results}
\label{sect:global_results}
The results from the \lya LF in each redshift interval are integrated with luminosity to map out the evolution of \lya luminosity density with redshift:
\begin{equation}
\rho_{Ly\alpha}= \int L_{Ly\alpha} \,\phi_{Ly\alpha} \, dL_{Ly\alpha}. 
\end{equation}
This is then converted to the \lya SFRD based on the calibration of \cite{kennicutt1998Schmidtlaw}, assuming the factor of 8.7 between intrinsic \lya and H$\alpha$ flux and a case B recombination \citep{Osterbrock_1989}:
\begin{equation}
\mathrm{SFRD_{Ly\alpha} [M_{\odot} yr^{-1} Mpc^{-3}] = 7.9 \times 10^{-42} \times \rho_{Ly\alpha} / 8.7}.  
\end{equation}

\indent The UV luminosity density is converted into the equivalent SFRD using the form $\mathrm{SFRD} = \kappa \times \rho_{UV}$ \citep{Madau_2014}, in which $\kappa$ is the conversion factor, sensitive to the recent SFH, metal-enrichment history and the IMF. The term $\rho_{UV}$ is the UV luminosity density, which is expressed in units of $\mathrm{erg\,s^{-1} Hz^{-1}}$, estimated by integration the UV LF with $\mathrm{M_{UV}}$. The widely used value of the conversion factor from \cite{kennicutt1998SFHubble} is $\kappa = 1.4 \times 10^{-28}$ , where a Salpeter IMF \citep{salpeter1955IMF} is assumed. \cite{Bouwens2022UVLF2z9} used a Chabrier IMF \citep{Chabrier_2003} for this conversion, so we correct for this, using a factor of 1.96. Going forward, the SFRD obtained from UV LF is treated as the total SFRD of the entire galaxy population. We estimated the total UV SFRD down to $\text{M}_{\text{UV}}=-13$ at redshifts $z = 2, 3, 4, 5, 6, 7$  (see Table~\ref{tab:UV_SFRD} for the values used, after correction to the Salpeter IMF). Subsequently we interpolated the SFRD to redshifts $z=2.5, 3.1, 3.5, 3.7, 4.5, 4.7, 5.4, 5.7$ to adapt to redshift bins used in the literature studies.\\

\begin{figure*}
  \centering
    \includegraphics[width=0.8\linewidth]{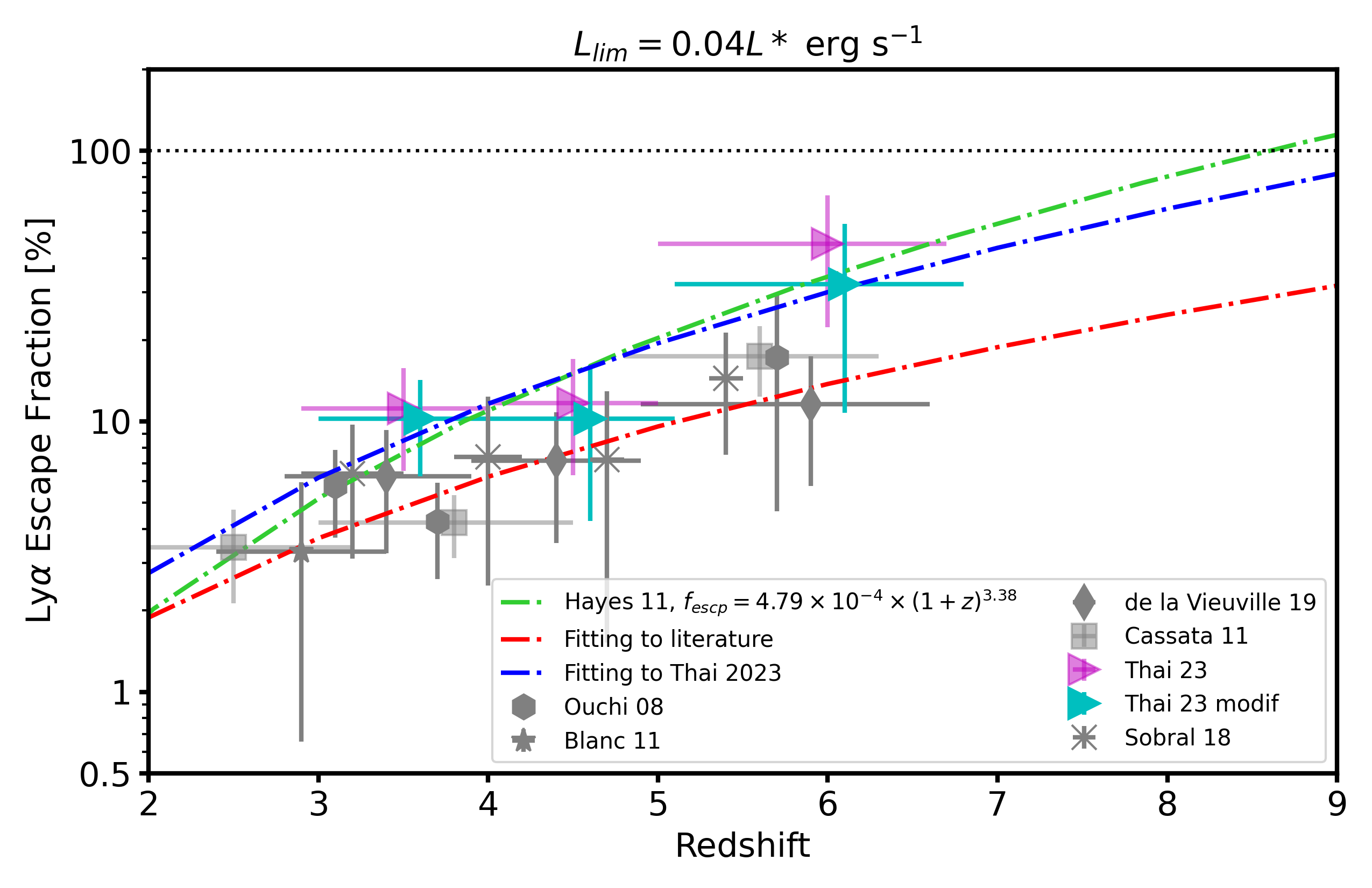}
\caption{Global redshift evolution of \flya with an integration limit on both LFs of $0.04L*$. Our results, using the LFs from \cite{Thai2023} are shown in cyan for the modified Schechter function and pink for the general Schechter function. The dotted horizontal line denotes $f_{esc} = 100\%$. The dash-dotted green curve is from \cite{hayes2011redshiftevofdust} in the redshift range of $2.2-6.6$ and calculated with the integration limit $0.04L*$. The red dash-dotted curve is the best fit to all the data shown and the blue dash-dotted curve fits just the data of \cite{Thai2023} (cyan points: modified Schechter function).}
    \label{fig: global escape fraction 0.04L*}
\end{figure*}

\indent The ratio between the LAE SFRD and the total, dust-corrected SFRD in a given redshift interval gives the global \lya escape fraction. This definition was applied in the works of \cite{hayes2011redshiftevofdust}, \cite{cassata2011vimosLAELF}, \citetalias{blanc2011}, \cite{Zheng_2013} and \cite{Konno_2016}. The results for the global escape fraction derived in this work are shown in Figs~\ref{fig: global escape fraction 0.04L*} and \ref{fig: global escape fraction 0.0006L*}. We compared our results to the results derived using several LAE LFs in the literature, for which we adopted an identical approach to that described above \citep{Ouchi2008LAE_LF, blanc2011, cassata2011vimosLAELF, GdlV2019LAELF, Thai2023, Sobral_2018}. We used the same UV LF for each comparison. \\
\indent The lower limits of the respective LF integrations are worth noting. We used two different limits, the first was $0.04L*$  and the second was $0.0006L*$ (where $L*$ is the characteristic galaxy luminosity of the Schechter function). The first limit, as suggested in \cite{hayes2011redshiftevofdust} is designed to compare our study to previous results using brighter galaxies. It corresponds to limits of $\mathrm{log\,L_{Ly\alpha} [\text{erg} \, \text{s}^{-1}]\sim41}$ and $\mathrm{M_{UV}}\sim-17$ respectively. The results for \flya calculated with this limit are shown in Fig.~\ref{fig: global escape fraction 0.04L*}. The second limit, $0.0006L*$, corresponds to the faint regimes that we reach with the LAE and UV LFs ($\mathrm{log\,L_{Ly\alpha}[\text{erg} \, \text{s}^{-1} ]\sim39.5}$ and $\mathrm{M_{UV}}\sim-13$) and is shown in Fig.~\ref{fig: global escape fraction 0.0006L*}. The vertical error bars are estimated from published uncertainty values in $\phi*$, $\alpha$, $L*$ (LAE LFs) and $\phi*$, $\alpha$, $M*$ (UV LF). The horizontal error bars display the redshift intervals of each survey.  As \cite{GdlV2019LAELF} and \cite{Thai2023} probed the LAE LF at the same redshift intervals, the data points derived from the LF of \cite{GdlV2019LAELF} have been shifted by 0.1 in redshift. Table~\ref{tab: LAE luminosity density} presents the results for each study using this integration limit, as well as the SFRD value derived. By comparing the results obtained with the different limits we can assess the impact of the faint galaxies that we reach with the latest LAE and UV LFs.   \\
\indent To investigate the global evolution of escape fraction with redshift, we recall the power-law form that was proposed by \citetalias{blanc2011} and \cite{hayes2011redshiftevofdust}: $f_{escp}^{Ly\alpha} = C \times \, \, (1+z)^\xi$. The coefficient $C$ is constrained strongly by the evolution of escape fraction at low redshift while the exponent $\xi$ is defined by the evolution at high redshift. In Fig.~\ref{fig: global escape fraction 0.04L*}, the best fits calculated for this work are shown as red and blue dash-dotted lines, fitting all the points in the graph and only those of \cite{Thai2023}, respectively. In Fig.~\ref{fig: global escape fraction 0.0006L*}, the red fit is to all the data. The green dashed line shows the fit from \cite{hayes2011redshiftevofdust}, which used the UV LF obtained from \cite{Ouchi_2004, Reddy2008LFs, Bouwens2009UV_LF_SFRD} and an LF integration down to $0.04L*$ (fitted to data over the range $2<z<6$). We leave this fit in both plots to guide the eye as to the difference when integrating to a fainter limit. \citetalias{blanc2011} use the results of \cite{Bouwens2010UV_LFs_z8gals} and an integration limit of 0. The fit of \citetalias{blanc2011} is shown in Fig.~\ref{fig: global escape fraction 0.0006L*} to provide the comparison to earlier results which have been integrated to 0.\\
\indent We find $C=14.3\pm14.0 \times 10^{-4}$ and $\xi=2.3\pm0.6$ when fitting to all the literature and $C=12\pm43 \times 10^{-4}$ and $\xi=2.8\pm2.1$ when only fitting the results from \cite{Thai2023} with the integration limit at $0.04L*$. With the lower integration limit of $0.0006L*$, we find $C=4.1\pm 7.7 \times 10^{-4}$ and $\xi=3.6\pm 1.2$ when fitting our results together with all the ancillary data in Fig.~\ref{fig: global escape fraction 0.0006L*}. Because the other studies included in this graph do not probe the faint regime that our LFs do (equivalent to the integration limit of $0.0006L*$), and they therefore require an extrapolation outside the limits of the datasets used. During this fitting process, we set the weight of each point equal to the weight of the equivalent point from our study (i.e. the cyan point in the same redshift bin). This relation is qualitatively consistent with the equivalent relation derived in \cite{hayes2011redshiftevofdust} (green dashed line, noting that this relation is derived using a different UV LF and a brighter integration limit) and suggests that \flya will reach 100\% around $z=8$. \\
 
\begin{table*}[]
    \centering
    \caption{\centering Ultraviolet SFRD used to calculate \flya.}
    \vspace{-0.2cm}
    \begin{tabular}{c c c c}
        \hline
        \hline
        \noalign{\smallskip}
        Redshift  & SFRD unobscured  & SFRD obscured & SFRD total   \\
        \hline
        \noalign{\smallskip}
        && $\mathrm{log_{10} [M_{\odot}\,Mpc^{-3}\,yr^{-1}}]$& \\
        \noalign{\smallskip}
        \hline
        \noalign{\smallskip}
         $z = 2.0$ & $-1.28\pm0.03$ & $-0.8\pm0.1$ &  $0.220\pm0.043$\\
         $z = 3.0$ & $-1.11\pm0.07$ & $-1.1\pm0.1$ & $0.163\pm0.026$\\
         $z = 4.0$ & $-1.12\pm0.12$ & $-1.2\pm0.1$ & $0.189\pm0.031$\\
         $z = 5.0$ & $-1.33\pm0.09$ & $-1.5\pm0.1$ & $0.078\pm0.013$ \\
         $z = 6.0$ & $-1.46\pm0.05$ & $-2.0\pm0.1$ & $0.045\pm0.005$\\
         $z = 7.0 $ & $-1.56\pm0.06$ & $-2.6\pm 0.1$ & $0.030\pm0.004$\\
         \hline
         \hline
    \end{tabular}
    \vspace{0.1cm} 
       \begin{tablenotes}
         \centering
        \small
         \item \textbf{Notes.} Values adapted from \citep{Bouwens2022UVLF2z9}, adjusted for the conversion from the  Chabrier IMF to the Salpeter IMF. Each value was calculated with an integration limit of $\mathrm{M_{UV}=-13}$.  
     \end{tablenotes}
    
    \label{tab:UV_SFRD}
\end{table*}

\section{Discussion}
\label{sect:Discussion}

With the individual and global results now in hand, we can compare them and consider what these can tell us about \flya and its evolution.

\subsection{Global \flya evolution}
\label{sect:global_fesc_discussion}

\begin{figure*}
    \centering
    \includegraphics[width=0.8\linewidth]{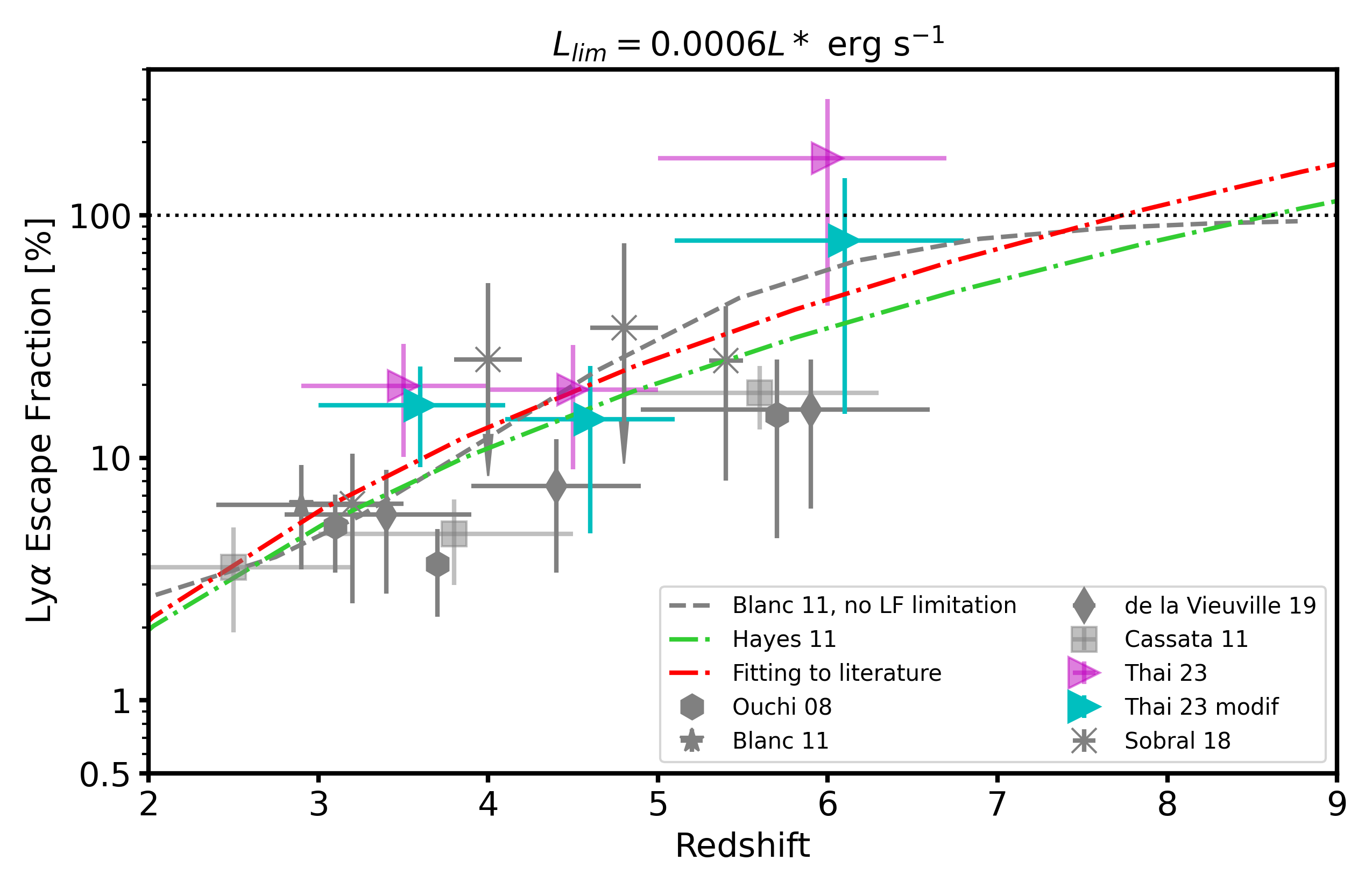}
    \caption{Global redshift evolution of \flya with an integration limit on both LFs of $0.0006L*$. Results from this study are shown by cyan (modified Schechter function) and pink (general Schechter function) points. Power-law fits from the literature are shown in grey and green (\citetalias{blanc2011}, \citealt{hayes2011redshiftevofdust}) and the equivalent fit to all the data is shown in red. The fit from \cite{hayes2011redshiftevofdust} is derived from results integrated to $0.04L*$, and the fit from \citetalias{blanc2011} from results integrated to 0.}
\label{fig: global escape fraction 0.0006L*}
\end{figure*}

Fig.~\ref{fig: global escape fraction 0.04L*} shows the determination of \flya using the LF integration limit $0.04L*$. We plotted this graph in order to compare our results to previous literature determinations of the escape fraction, which used samples of galaxies brighter than this work, and to assess the impact of the bright ($\mathrm{M_{UV}<-17}$) and faint populations ($\mathrm{M_{UV}<-13}$) on the evolution of \flya. We plotted results derived using both the modified (cyan) and the general Schechter (pink) functions in order to show the difference in these results created by the presence or lack of a turnover in the LAE LF. However, at this integration limit, the behaviour of the modified Schechter function is close to that of the traditional Schechter function, so we did not see a significant difference in $\text{SFRD}_{Ly\alpha}$. As the current best derivations of the LF come from the modified Schechter functions, going forth we discuss the results related to these (cyan points).\\
\indent We observed a good agreement at all redshifts between the results derived in this work and the results from the literature. We saw little evolution between $z\sim3.5$ and $z\sim4.5$, followed by a jump in \flya at $z\sim6$, but at each redshift, our results agree well with the established literature. When fitting our data (blue line) compared to all the results (red line), we observed an enhancement of \flya, likely due to the constraints placed by the literature in the redshift range $2<z<4$. Our points do lie systematically above the points from the literature, indicating a greater SFRD based on the latest LAE LF of \cite{Thai2023}. We note, however, that the LFs of \cite{Konno_2016} and \cite{drake2017MUSELAELF} have not been included in Figs~\ref{fig: global escape fraction 0.04L*} and \ref{fig: global escape fraction 0.0006L*} due to large uncertainties, especially at the highest redshifts and faint luminosities. These uncertainties contribute to unrealistically high (and poorly constrained) escape fractions based on these LFs.\\
\indent Additionally, the choice of completeness cut has an impact. In \cite{Thai2023}, a completeness cut of $1\%$ was chosen (i.e. LAEs with completeness values $<1\%$ are not taken into account in the LF determination), in order to maximise the sample size, while still negating the effect of the most incomplete and therefore uncertain sources. Studies have typically chosen different completeness cuts, or none at all. For example, \cite{GdlV2019LAELF} chose $10\%$, rejecting a larger number of sources. In order to check the effect this has, we recomputed \flya using the LF cut at 10\% completeness. This yielded values similar to those seen in the literature: $10-15\%$ escape. Therefore, we consider our results completely consistent with what has been found previously when considering a brighter integration limit ($0.04L*$).\\

\indent The green dashed fit of \cite{hayes2011redshiftevofdust} lies slightly higher than the literature data. We attribute this to the significant update in the UV LF since the publishing of that study (the UV LFs of \cite{Bouwens2009UV_LF_SFRD} and \cite{Reddy2008LFs} were used for this fit). This emphasises the importance of the UV LF in these calculations and the value of the latest determination from \cite{Bouwens2022UVLF2z9}. The values used in this work, inclusive of dust and IMF correction are shown in Table~\ref{tab:UV_SFRD}. \\
\indent The values we derive for \flya are considerably higher when we extend the LF integrations down to $0.0006L*$, particularly for the highest redshift bin, $z\sim6$, where we see an increase of 0.4 dex with respect to the $0.04L*$ integration. In fact, this point is consistent with the fit to the data in \citetalias{blanc2011}, which is integrated to zero, that is, taking into account the full population by assuming the validity of the two LFs down to zero. This large escape fraction value is unsurprising, as we see a high number density of faint galaxies in the LAE LF, and these populations typically have higher \flya (see Fig.~\ref{fig:MUV_fesc}, as well as \citealt{goovaerts2023,Chen2023Lya_escape_JWST,Goovaerts2024,Napolitano2024lya_visibility_reion,nakane2024high_z_LAEs}). In the $z\sim6$ bin, \flya is consistent with $100\%$, i.e. a similar SFRD derived from LAEs as from the UV-selected population. While still consistent at $1\sigma$ with the fit derived by \citetalias{blanc2011} (using an LF integration to 0), our point lies above points from the literature. We note that no other studies have had access to the faint regime of LAEs used in this study and as such, the literature LFs we compare to are integrated beyond the limits of their samples to create this graph. Reasons for this enhanced \flya value for faint LAEs at $z\sim6$ include the effect of the LAE incompleteness treatment, which acts most strongly for faint objects at the highest redshifts. Effectively, this would mean that we are counting in the LAE LF a significant number of faint LAEs not seen in the continuum.\\
\indent This result may also point to a general scenario where faint galaxies (i.e. $\mathrm{M_{UV}\sim-13}$) generally see an extremely high production and escape fraction of \lya photons (see also \citealt{Atek2024dwarf_gals_reion}). This appears to be a clear effect when integrating down to the $L*$ limit our sample reaches: faint galaxies experience a higher escape of \lya photons, especially at higher redshifts. This result is also in agreement with the results presented in Fig.~\ref{fig:MUV_fesc} as in this plot, between $\mathrm{M_{UV}\sim-17}$ and $\mathrm{M_{UV}\sim-13}$, we see LAEs with exclusively high escape fractions (almost all above 10\% and many around 100\%). This result implies that at $z\sim6$ and beyond, faint galaxies tend to be highly star forming and chemically unevolved, which should allow high escape fractions. On the other hand, seen in the uncertainties in the faint end of the LAE LF, it is clear that there is still a large intrinsic scatter in the populations (once more in agreement with the results in Section~\ref{sect:individual_escape}). \\
\indent It is worth noting, however, that if the IMF of these faint, reionisation-era LAEs is top-heavy with respect to the assumed Salpeter IMF, this may also explain an elevated \flya. Such an IMF would increase the number of ionising photons, from massive stars, which would in turn increase the \lya emission of the galaxy. A higher ionising photon production efficiency, $\xi_{ion}$ has been observed for some LAEs (\citealt{matthee2021xshooterLAEs,Saxena2024xi_ion_LAEs}, see also \citealt{Simmonds2023xi_ion_LAEs_z6}) although there is no firm consensus on how much higher. If $\xi_{ion}$ were indeed to be $25.5-25.6$, this would mean a factor of $2-3$ more ionising photons produced. This would in turn mean more \lya photons produced and an overestimation of \flya in our results by a similar factor (we note that this would apply to all the results in Figs.~\ref{fig: global escape fraction 0.04L*} and \ref{fig: global escape fraction 0.0006L*}). This is an important dependance on the high-mass slope of the IMF and the production of ionising photons, and could be particularly relevant to the faintest galaxies towards the epoch of reionisation, which are more likely to host exotic stellar populations. \\ 
\indent In summary, we find that the LAE SFRD matches the UV SFRD during the final stages of reionisation, and is $\sim20\%$ of it at redshifts between 3 and 5. For both bright and faint integration limits, we see this jump in the highest redshift bin ($z\sim6$) compared to little evolution between redshifts of 3 and 5. This reflects a significant evolution of the LAE LF in this redshift bin, therefore suggesting that the LAE population evolves significantly between the end of reionisation ($z\sim6$) and later times ($z\sim4.5$). Although it is worth noting the uncertainty on our $z\sim6$ result that we attribute to a combination of several factors: a lower dust content in galaxies at this redshift and above combined with higher specific SFRs in our sample of faint galaxies. As mentioned, the completeness correction for the LAE LF acts most strongly on the faint, high-redshift LAEs, making this a possible additional contributor. 

\subsection{Comparison of the global and individual \flya evolutions}
\label{sect:discussion_comparison}

It is immediately clear that the global evolution of \flya shows a more significant trend than the individual evolution, despite significant uncertainties, especially in the highest redshift bin. This becomes additionally clear when incorporating the agreement in the literature for the global evolution.\\
\indent For our datasets, this is likely due at least in part to the far greater sample size used in the global evolution: 2500 UV-selected galaxies and 600 LAEs, compared to $\sim100$ LAEs for which \flya is calculated (Fig.~\ref{fig:Fesc_individual}). For a smaller sample such as the one used for the individual tracking of \flya, particularly considering the lack of sufficient galaxies at redshifts greater than 6, any redshift evolution is washed out by the scatter in individual results. The results of \citetalias{blanc2011} display a similar lack of evolution over $1.9<z<3.8$ when taking into account the significant scatter of individual galaxies. The same is even true for the more recent studies of \cite{Chen2023Lya_escape_JWST} and \cite{Napolitano2024lya_visibility_reion}, using Balmer decrements to calculate \flya, albeit with smaller sample sizes. The intrinsic variety of escape scenarios for \lya photons from different galaxies (even bearing in mind that all these systems are selected as LAEs) is evident. \\
\indent The trend seen in the global evolution of \flya (integration limit $0.0006L*$), towards \flya$\sim80\%$ around $z=6$, is not seen in the individual results. Likely this is at least in part due to the incompleteness of the sample used to calculate \flya for individual galaxies at this redshift, as both LAEs with very small escape fractions and LAEs with very large escape fractions are harder to observe. For the LAEs with large escape fractions this difficulty comes from the faintness of the associated continuum: the galaxies with the highest \flya values tend to be faint in terms of UV magnitude (Fig.~\ref{fig:MUV_fesc}). This incompleteness should be accounted for in the two LFs used, so we do not expect this to play a significant role for the global \flya evolution.\\
\indent Contrasting these two results, global and individual, it is clear to see that a large sample size, as well as considered completeness corrections, are necessary to observe any significant redshift evolution of \flya. Indeed, further exploration of the faintest regimes of LAEs (and UV-selected galaxies) would be desirable in order to better constrain the faint end of the LFs and reduce the uncertainties on the highest-redshift points. For example, it is still very challenging to observe more than a few galaxies with $\mathrm{log\,L_{Ly\alpha}<40}$ at $z>5.5$ (see, for example, Fig.~3 of \citealt{Thai2023}).\\
\indent We note also that the global comparison is effectively the \lya escape fraction of a given volume of the Universe, bounded by the luminosity and redshift limits of the surveys used. This calculation takes into account all galaxies, not just those actually selected as LAEs, as is the case for the individual results. We might therefore reasonably expect to see a smaller escape fraction due to the inclusion of galaxies not selected as LAEs. In the lower and intermediate redshift bins, the results are consistent with the sample median shown by the red dashed line in Fig.~\ref{fig:Fesc_individual}; however in the highest redshift bin, the global calculation of \flya appears to yield a larger value. Taking the median of the galaxies from Fig.~\ref{fig:Fesc_individual} above $z=5$ would yield a higher result (36\% escape), and it is likely that we simply do not detect enough LAEs of high escape fraction at higher redshifts in our limited sample behind A2744. As this is just one cluster, extending these measurements to other clusters efficient in lensing galaxies above $z=5$ is very desirable. \\
\indent The galaxies detected by their \lya emission but not in the continuum (the black triangles in Fig.~\ref{fig:MUV_fesc}) naturally all have high escape fractions and are counted in the LAE LF, so this may go some way towards explaining the discrepancy we observe between our different methods of calculating \flya. Additionally, and as mentioned, the completeness correction added to the LAE LF acts strongly in the faint, high-redshift regime so we also expect this to play a role. The results in Fig.~\ref{fig: global escape fraction 0.0006L*} suggest that this may be a stronger role than the equivalent completeness correction added to the UV LF. \\
\indent Contrary to the lack of similarity in redshift evolution in the global and individual \flya results, there is a similar trend in terms of an evolution with the UV magnitude of the host galaxy. In Sect.\ref{sect:individual_escape} this is established as a much more significant trend and the results when comparing LFs support this. When integrating to fainter limits we see significantly higher values of \flya across the redshift range probed. 

\subsection{The overlap of the LAE and ultraviolet luminosity functions}
\label{sect:undetected_LAEs}

\begin{figure}
    \centering
    \includegraphics[width=0.5\textwidth]{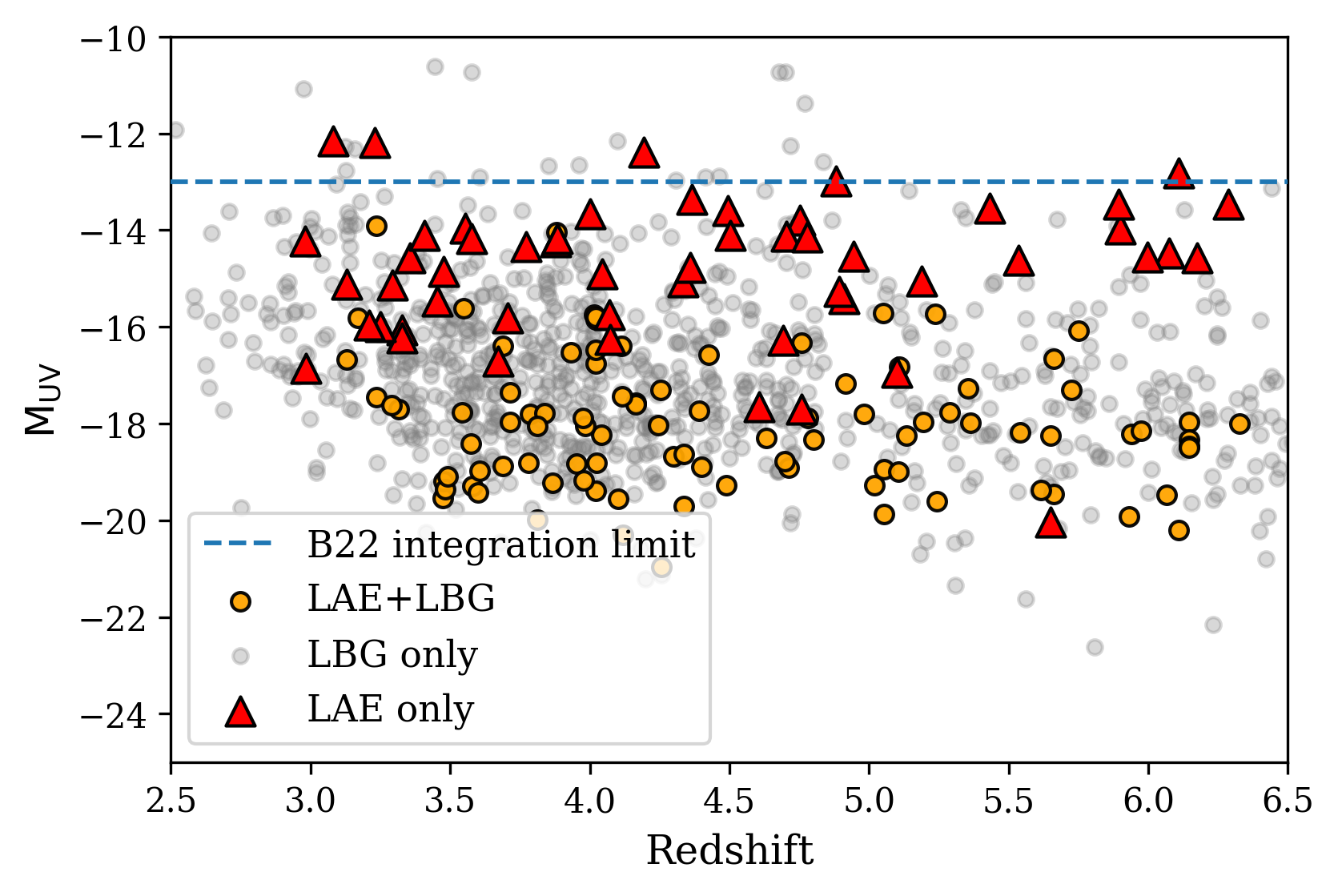}
    \caption{Redshift and $\mathrm{M_{UV}}$ distribution of LAEs and Lyman-break galaxies (UV-selected galaxies) from the sample in \cite{goovaerts2023}. Grey points denote the Lyman-break galaxies and are thus analogous to the UV-selected sample. Orange points denote objects that were selected as LBGs and LAEs and upwards-facing triangles indicate LAEs that were not detected at all in the HST imaging \citep{shipley2018hff}. The $2\sigma$ upper limits for the continuum of these objects have been estimated using the local noise in the region where they would have been detected in the HST image that would see the $1500\,\AA$ emission. The horizontal dashed line denotes the integration limit in the \rtot calculation in \cite{Bouwens2022UVLF2z9}.}
    \label{fig:LAEonly}
\end{figure}

Here we revisit the assumption made in order to compare the two LFs using single integration limits in terms of $L*$ and treat them as representative of their respective ``total'' populations. This, in principle, is troublesome as there is no direct mapping between \lya luminosity and UV magnitude; for every given \lya luminosity bin, there are a range of UV magnitudes that galaxies display. In fact, this encodes exactly what we want to measure, the escape of \lya photons. It is clearly invalid to compare a LF based on just the brightest LAEs to the complete UV LF, including the faintest sources. Therefore, in order to make the comparison discussed above and shown in Figs.~\ref{fig: global escape fraction 0.04L*} and \ref{fig: global escape fraction 0.0006L*}, we have to assume that we adequately represent the ``total'' populations with both LFs, within the UV and \lya luminosity integration limits used. This approach is discussed at length in \cite{hayes2011redshiftevofdust} and \citetalias{blanc2011} and we revisit this in the context of our dataset.\\
\indent That we adequately represent both \lya and UV populations is reasonably evident for the brightest objects, they are certain to be included in their respective surveys. This is less evident for the fainter regimes explored. For example, it is useful to ask whether the faintest LAEs included in the LAE LF are adequately represented by the faint end of the UV luminosity function. In order to assess this, we can look at the faintest LAEs and compare them to the UV magnitude regimes included in the UV LF of \cite{Bouwens2022UVLF2z9}.\\
\indent It is by now well established that certain LAEs remain undetected in even the deepest photometric searches to date (see, for example, \citealt{bacon2015muse_details,Maseda2018MUSE_faintELGs,sobral2019predictinglyaesc,Maseda2020faintLAEs,GdlV2020LAEfrac,goovaerts2023}). Additionally, some LAEs display faint continuum that would not pass the criteria for many UV-based selections, be they based on colour-colour cuts \citep{stark2010keckLAEfrac,pentericci2011laefrac/z=7LBG,pentericci2018LAEfrac} or photometric redshifts \citep{caruana2018,kusakabe2020,goovaerts2023}. We note that preliminary searches conducted by JWST/NIRCam in lensing clusters yields almost exactly the same number of undetected LAEs \citep{Goovaerts2024}. On the other hand, nebular lines are detected in these systems, so they are likely to be real detections \citep{Maseda2023UVdarkLAEs}. With future, deep JWST (and HST) surveys, we will be able to shed more light on this issue.\\
\indent This raises concerns as to whether the UV LF truly represents the entirety of the galaxy population, or whether there are a significant number of `UV-dark' LAEs that are counted in the LAE LF but not in the UV LF. This is important to consider given the high values of \flya we derive when integrating to $0.0006L*$ (Fig.~\ref{fig: global escape fraction 0.0006L*}).\\
\indent A way that we can assess which of these two cases is closer to reality is to compare the UV magnitude regime where these HST-dark LAEs are expected to lie to the regime covered by the UV LF in \cite{Bouwens2022UVLF2z9} and the integration limits we use. Upper limits on the UV magnitude of these LAEs can be calculated by manually extracting the continuum level from the HST images, as in \cite{goovaerts2023}. These objects, denoted by upward-pointing triangles to indicate upper estimates on $\mathrm{M_{UV}}$, are displayed in Fig.~\ref{fig:LAEonly}. This shows that for the clusters considered by the UV LF in question, and for the photometry in question, the UV-dark LAEs would be expected to appear around $\mathrm{M_{UV}}=-16$, with many residing between this and $\mathrm{M_{UV}}=-14$, and some down to $\mathrm{M_{UV}}\sim-12$. \\
\indent The UV LF in \cite{Bouwens2022UVLF2z9}, inclusive of the correction for selection incompleteness, extends past $\mathrm{M_{UV}}=-16$ at all redshifts relevant for this work, and extends past $\mathrm{M_{UV}}=-14$ at $z\sim3$ and $z\sim6$, and close to it at $z\sim4$ and $z\sim5$. At $z\sim3$, the coverage extends all the way to $\mathrm{M_{UV}}=-12$. The integration limit ($0.0006L*$) used for the calculation of \rtot from the UV LF corresponds to $\mathrm{M_{UV}}=-13$, as denoted by the horizontal dashed line in Fig.~\ref{fig:LAEonly}.\\
\indent There are four LAEs estimated to have upper UV magnitude limits fainter than this in the sample of four lensing clusters in \cite{goovaerts2023}. As the non-detection of these LAEs likely depends at least in part on the relative depths of the spectroscopy and photometry, it is difficult to estimate what effect this has across our sample, but we can reasonably expect this to have a minor effect, considering the small number of LAEs in this position in comparison to the sample size used for the two LFs. \\
\indent We can additionally reassure ourselves by comparing what the UV luminosities should be from the faintest LAEs assuming different levels of \lya escape. The faintest LAEs included in the LAE LF are no fainter than $\mathrm{L_{Ly\alpha}}=10^{39}\,\mathrm{erg\,s^{-1}}$, which equates to an absolute UV magnitude of $\mathrm{M_{UV}}\sim-10.4$ assuming 100\% \lya escape (meaning this value is a lower (faint) limit on the absolute UV magnitude). If we assume a more likely \flya value of our sample median from Section~\ref{sect:individual_escape} of 22.5\%, this UV magnitude becomes $\mathrm{M_{UV}}=-12.2$. This is very faint compared to the reach of the UV LF we use, below the integration limit of $\mathrm{M_{UV}}=-13$. However, using this integration limit and an escape fraction of 22.5\%, we recover a value $\mathrm{L_{Ly\alpha}}=2.4\times10^{39}\,\mathrm{erg\,s^{-1}}$. There is only one LAE in our sample close to this luminosity (see Figure 3 of \citealt{Thai2023}). On the other hand, following this analysis shows that faint LAEs with $\mathrm{L_{Ly\alpha}}<10^{40}\,\mathrm{erg\,s^{-1}}$ are not accounted for in the UV LF with its integration limit set at $0.0006L*$ as in Fig.~\ref{fig: global escape fraction 0.0006L*} if they have \flya$>95\%$. Across the board, this effect will bias our results for \flya low.\\
\indent We can perform the same exercise in an inverse manner, to assess whether we are likely to see the \lya emission from the faintest UV objects included in the UV LF. The LAE luminosity expected from a galaxy of $\mathrm{M_{UV}=-12}$ is $10^{39.6}\,\mathrm{erg\,s^{-1}}$, assuming 100\% \lya escape. Assuming 22.5\% escape, this value will be $10^{39.0}\,\mathrm{erg\,s^{-1}}$. So for galaxies of this absolute magnitude, we can expect to account for all the LAEs in the population (when taking into account the completeness correction). However the escape fraction can be lower (down to 0\%), meaning that we will miss the objects of very low UV magnitude and very low \lya escape. Unlike missing objects in the UV selection, this biases our results towards high \lya escape values. This counter-bias can alleviate the bias introduced by missing UV-faint objects as described in the previous paragraph, and indeed this is the motivation behind choosing one value of $\mathrm{L_{\star}}$ down to which the two LF integrations are calculated.\\
\indent In summary, and in light of the high values for \flya found in Fig.~\ref{fig: global escape fraction 0.0006L*}, for these results to be overestimated, a large population of continuum-undetected LAEs must be playing a role (together with the effects introduced by their incompleteness) in the LAE LF that are not accounted for in the UV LF, even with the completeness treatment described in \cite{Bouwens2022UVLF2z9}. This potential effect is important for studies looking to evaluate the total SFRD of the Universe. \\
\indent Alternatively, if we assume that this effect is negligible and the high \flya value derived at $z\sim6$ when including our whole faint sample is reflecting a real similarity between $\mathrm{SFRD_{Ly\alpha}}$ and $\mathrm{SFRD_{UV}}$, this indicates that galaxies at $z\gtrsim6$ are emitting significantly higher amounts of \lya photons than later in the Universe. This supports a picture of galaxies towards reionisation as chemically unevolved and highly star forming, creating a lot of \lya emission and allowing it to escape.

\subsection{Revisiting the contribution of LAEs to cosmic reionisation}
\label{sect:LAE_contribution}

\begin{table*}
    \centering
    \caption{\centering LAE contribution to reionisation.}
    \begin{tabular}{ m{2.8cm} m{2.8cm} m{1.3cm} m{3cm} m{3cm} }
        \hline
        \hline 
        \vspace{1mm}
         Key & $\rho_{Ly\alpha}$ $(10^{39})[\mathrm{erg\,s^{-1}\,Mpc^{-3}}]$ & \flya [$\%$] & Contribution [$\%$] (\flyC$=0.4\%$) & Contribution [$\%$] (\flyC$=10\%$) \\

        \hline
        \vspace{1mm}
        \citetalias{Thai2023}, median & $47.31\pm37.63$ & 22.5 & $9^{+16}_{-7}$ & $260^{+210}_{-50}$\\
        \hline
        \vspace{1mm}
        \citetalias{Thai2023}, $z>5$ & $47.31\pm37.63$ & 36 & $6^{+10}_{-4}$ & $160^{+130}_{-30}$\\
        \hline
        \vspace{1mm}
        \citetalias{GdlV2019LAELF}, median & $9.52\pm5.65$ & 22.5 & $1.9^{+1.1}_{-0.8}$ & $52^{+31}_{-21}$\\
        \hline
        \vspace{1mm}
        \citetalias{GdlV2019LAELF}, $z>5$ & $9.52\pm5.65$ & 36 & $1.2^{+0.7}_{-0.4}$ & $52^{+31}_{-21}$\\
        \hline 
        \vspace{1mm}
        \citetalias{Thai2023} GSF, median & $103.2\pm76.5$ & 22.5 & $20^{+15}_{-5}$ & $560^{+420}_{-150}$\\
        \hline
    \end{tabular}
    \vspace{0.1cm}
    
    \begin{tablenotes}
      \small
      \item \textbf{Notes.} The key indicates the provenance of the values used in columns two and three, in the order they appear in the table. GSF indicates where the general Schechter function was used rather than the modified Schechter function. Values of $\rho_{Ly\alpha}$ come from the LAE LF integrated to $0.0006L*$ ($\mathrm{log\,L_{Ly\alpha}\sim39}$). \citetalias{Thai2023}, \citetalias{GdlV2019LAELF} refer to \cite{Thai2023} and \cite{GdlV2019LAELF} respectively. 
    \end{tablenotes}
    \label{tab:LAEcontribution}
\end{table*}

Seeing as the escape of \lya photons is a crucial quantity for the assessment of the LAE contribution to cosmic reionisation, we can now revisit this in light of our results. There is significant debate in the literature about the relative contribution of the LAE population to reionisation, (see, for example \citealt{cassata2011vimosLAELF,drake2017MUSELAELF,GdlV2019LAELF,matthee2022brightLAEs,Thai2023}). All of these results are, naturally, dependent on the \flya values assumed in the various studies. The \flya value used to calculate this contribution is often pivotal in the classification of this contribution. We can now recalculate the LAE contribution taking into account our results.  \\
\indent We started by considering the ionising emissivity from SFGs, using the following formalism \citep[e.g.][]{Madau_1999, robertson2013reionisation, Duncan_2015}: 
\begin{equation}
    \label{eq:ionising_emissivity}
    \dot{n}_{ion}(z) = \rho_{UV}(z) \, \xi_{ion}  \,\flyC,
\end{equation}
where $\dot{n}_{ion}$ is expressed as the product of the UV luminosity density $\rho_{UV}(z)$ \citep{Bouwens2015planckreionisation, Finkelstein_2016, Oesch_2018}, global UV luminosity to ionising photon conversion factor $\xi_{ion}$ \citep{Matthee_2017a, Lam_2019}, and the ionising photon escape fraction \flyC. 
We reformulated Eq.~\ref{eq:ionising_emissivity}, as we are interested in the quantity $\dot{n}_{ion}(z)$ for the LAE population only: 
\begin{equation}
    \dot{n}_{ion, Ly\alpha}(z) = \rho_{Ly\alpha}(z)  \, \xi_{ion}^{Ly\alpha}  \, \flyC,
\end{equation}
where $\rho_{Ly\alpha}$ is the \lya luminosity density, $\xi_{ion}^{Ly\alpha}$ is the conversion factor of the observed \lya luminosity to the number of ionising photons, and \flyC is the Lyman continuum escape fraction of the LAEs. Ignoring collisionally excited emission and using Case B recombination (\citealt{Brocklehurst_1971, Osterbrock_2006}; see also \citealt{matthee2022brightLAEs}) as well as the intrinsic flux ratio between \lya and $H\alpha$ of 8.7, the ionising emissivity of LAEs, corrected for \flya, is written as:
\begin{equation}
    \label{eq:final_lya_emissivity}
    {\dot{n}_{ion, Ly\alpha}(z) = \frac{\rho_{Ly\alpha} (z) \, \times \flyC}{8.7 \, c_{H\alpha} \times \flya \, (1-\flyC)}},
\end{equation}
where $c_{H\alpha}$ is the coefficient factor of the $H\alpha$ emission line, $\sim 1.25-1.35 \times 10^{-12}$ for Case B recombination. It remains to estimate the individual parameters appearing in the above function. \\
\indent - The LAE luminosity density ($\rho_{Ly\alpha}$): this parameter was presented in Sect. \ref{sect:global_results}. \\
\indent - The escape fraction of \lya photons, \flya. We cannot correct $\rho_{Ly\alpha}$ by the escape fraction derived in Sect. \ref{sect:global_results} as this, by construction, corrects $\rho_{Ly\alpha}$ to the total value of the parent population ($\rho_{total}$) that we take from \cite{Bouwens2022UVLF2z9}. We can nonetheless use the value we have derived in Sect.~\ref{sect:individual_escape}. We discuss both our sample median, \flya$=22.5\%$ and the sample median for the LAEs at $z>5$, \flya$=36\%$. \\
\indent - LyC escape fraction, \flyC: This parameter is poorly understood, in particular at high redshift, due to the increasing IGM neutrality, which absorbs Lyman continuum photons along the light of sight. However, recent progress has been made in local galaxies (high-redshift analogues) \citep{Flury2022LyC_lowz_survey,Izotov2024LAEsLyC_local} and in LAEs up to $z\sim3$ \citep{Liu2023LAEs_LyC_z3.1,Kerutt2024LAE_LyC,Pahl2024lya_line_lyC_escape,Jung2024LyCesc_lowmassgals}. However, results from these studies, as well as simulations \citep{Choustikov2024Lya_LyC_escape}, suggest that there is a significant scatter in the \flyC across the LAE population. Therefore, placing one value on \flyC, such as \flya, is a simplification. Many observational studies point to values between 10\% and 15\%: \cite{Kerutt2024LAE_LyC} derive an underlying population value of $12\%$, \cite{Jung2024LyCesc_lowmassgals} obtain values between $3\%$ and $15\%$, \cite{Liu2023LAEs_LyC_z3.1} give an upper limit of $16\%$ and \cite{Izotov2024LAEsLyC_local} find values between $1\%$ and $34\%$, with most of their galaxies having $\flyC<15\%$. Based on the SPHINX simulations \citep{Garel_2021} and the model for Lyman continuum escape developed by \cite{Rosdahl_2022}, \flyC is $1.17\%, 0.4\%$ at $\mathrm{M_{UV}} = -17, -13$ (we note that this is for all galaxies, not just those selected as LAEs).\\
\indent The ability to turn Eq.~\ref{eq:final_lya_emissivity} into a summation, with each LAE contributing its own $\rho_{Ly\alpha}$, \flya and with \flyC consequently following a relation such as those presented in \cite{Rosdahl_2022}, \cite{Choustikov2024Lya_LyC_escape} or \cite{Izotov2024LAEsLyC_local} would be an improvement. Increased statistics, especially for faint LAEs and at redshifts greater than 5 would be necessary to make this approach feasible. This is within reach for JWST in the coming years, as observed samples increase in size and more faint galaxies are uncovered.\\
\indent A very recent approach by \cite{Choustikov2024Nion_evolution_JADES} attempts this individual approach for a large sample of JWST/NIRCam-observed galaxies (not just LAEs) (observed by the JADES programme, \citealt{Eisenstein2023JADES_I}). The authors conclude that faint galaxies ($\mathrm{M_{UV}\geq-18.5}$) are potentially responsible for $\sim80\%$ of the reionisation budget. The majority of our sample of LAEs fall within this bracket (see, e.g. Fig.~\ref{fig:LAEonly}).\\
\indent With $\dot{n}_{ion, Ly\alpha}$, at $z=6$, from Eq.~\ref{eq:final_lya_emissivity}, we can compare this value to the ionising emissivity required for reionisation as derived by \cite{bouwens2015UVLF}, $\dot{n}_{ion,crit}=50.92$ (we note that this value uses a fiducial clumping factor $C_{\mathrm{HII}}=3$). In Table~\ref{tab:LAEcontribution}, we show the range of results, applying different values for $\rho_{Ly\alpha}$, \flya and \flyC. For $\rho_{Ly\alpha}$ derived from our results, (the LF of \citealt{Thai2023}) we find LAE contributions ranging from minor (5-10\%) for a low \flyC ($0.4\%$) to dominant $>100\%$ for \flyC$=10\%$. Changing \flya between the sample median and $z>5$ median does not qualitatively change the outcome. As \flyC changes more significantly depending on the chosen value, this represents the main effect on the outcome. In several cases we see unrealistic contributions ($>100\%$) for LAEs based on \flyC$=10\%$, suggesting that this value may be too high and not representative of our whole sample. Crucial work remains to correlate \flyC with observable parameters (such as \lya parameters) at higher redshifts and constrain the escape of ionising photons for the LAE population at $z\sim6$ and above. \\
\indent We investigate the effect on the LAE reionisation contribution in the case of a completeness cut in the calculation of the LAE LF at 10\%. We take the results of \cite{GdlV2019LAELF} as emblematic of this scenario (as mentioned, the LF of \cite{Thai2023} is consistent with that of \cite{GdlV2019LAELF} if this cut is applied). In this case, the contribution to reionisation derived for the LAE population is significantly less ($1-2\%$ for \flyC$=0.4\%$ and $30-50\%$ for \flyC$=10\%$). This emphasises the sensitivity of these results to the completeness treatment applied for the LAE population.\\ 
\indent We additionally show the difference between the modified Schechter function, which we have chosen to adopt throughout this study, and the general Schechter function (noted GSF in Table~\ref{tab:LAEcontribution}. Using the general Schechter function roughly doubles the LAE contribution to reionisation. \\
Highlighting the effects of both the function chosen to fit the LAE LF and the completeness cut applied serves to reinforce the need for a larger sample at $z>5$ and for faint galaxies. Better statistics in this parameter space will constrain the faint end of the LF more strongly, discriminating between the modified and general Schechter functions. With deeper observations of faint galaxies, the completeness for the faint LAEs will also improve (i.e. the sample will be more complete at a given \lya luminosity) and the chosen cut will become a less important factor.\\ 
\indent The chosen value of \flyC is clearly determinant in this calculation with LAEs contributing very little for small values of \flyC and providing the dominant contribution for reasonable values of \flyC observed in many lower-redshift LAEs and low-mass galaxies acting as high-redshift analogs. Using the $\rho_{Ly\alpha}$ from \cite{Thai2023} and \flya$=36\%$, a sample average \flyC of just $6.5\%$ would be enough for LAEs to provide all the necessary ionising photons at $z=6$. If the values of \flyC observed in these low-redshift LAEs are maintained across the LAE population into reionisation, it is certainly conceivable that these galaxies could have been responsible for reionisation in its entirety. We note that the results in this study do not constrain the reionisation scenario beyond $z=6.7$ and that LAE LFs at $z\geq7$ are needed to track the contribution of LAEs further into the EoR.

\section{Conclusions}
\label{sect:conclusion}
We have derived the escape fraction of \lya photons across the redshift range $2.9<z<6.7$ using two methods, namely, comparing dust-corrected star formation to star formation derived from \lya emission and comparing the SFRD from the UV `parent' population LF and the \lya SFRD derived from the LAE LF. Here we summarise the main conclusions of the work.\\
\indent -- We determined the values of \flya for 96 individual galaxies behind the A2744 lensing cluster, finding a significant scatter and a sample median of $22.5\,\%$. \\
\indent -- We observed a significant evolution of \flya with absolute UV magnitude, following the relation $\mathrm{log(\flya)=(0.28\pm0.03)\,M_{UV}+(4.3\pm0.6)}$. Notably, we saw very few LAEs with \flya$<10\,\%$ among objects fainter than $\mathrm{M_{UV}=-18}$.\\
\indent -- We expect the line-incompleteness of MUSE \lya detections to play a role here, but only for the faintest objects, $\mathrm{M_{UV}\gtrsim-14}$.\\
\indent -- For the global redshift evolution of \flya, we derived results in agreement with previous determinations in the literature when using a bright LF integration limit ($0.04L*$). \flya appears to reach 100\% around $z=11-12$.\\
\indent -- When we assessed \flya with an integration limit that reflects the faint limits of our sample ($0.0006L*$), we found that \flya becomes consistent with 100\% escape by $z\sim6$. This result supports galaxies being chemically unevolved and highly star forming at these redshifts, producing a high quantity of \lya photons and allowing them to escape. \\
\indent -- With both bright and faint integration limits, we observed little evolution between $z=3-5$ and a jump at $z\sim6$, suggesting a rapid evolution of the LAE population towards reionisation, although this effect is at $<1\sigma$ for our dataset. \\
\indent -- We saw more significant redshift evolution in our global assessment than for individual galaxies due to the large intrinsic scatter in the LAE population combined with the far greater statistics in the respective LFs used for the global comparison.\\
\indent -- We revisited the LAE contribution to the reionisation process, taking into account our new values for \flya and a range of motivated values for \flyC. The results depend heavily on the assumed values for \flyC. Taking the latest values observed in LAEs at $z\sim3$, we find that LAEs can provide all the ionising emissivity needed for reionisation. \\
\indent -- We demonstrated the additional strong dependence of the LAE contribution on the completeness cuts and fitting form of the LAE LF. This stresses the urgent need for better statistics at $z>5$ and $\mathrm{log\,L_{Ly\alpha} [erg/s]<40}$.\\
\indent At the moment, it is challenging to improve the two luminosity functions used in this work, as they already use the full extent of deep observations in lensing clusters from HFF and MUSE. Further clusters to be included in this sample are desirable, particularly to better constrain the high-redshift ($z\sim6$) faint end of the LAE LF. This will be made possible through further MUSE-observed lensing clusters, such as those presented in \cite{Claeyssens2024LLAMAS2}. This will also help bin the results into smaller redshift bins to better constrain the precise evolution in the range $3<z<6$. Observations of LAEs at higher redshift from JWST, such as \cite{Chen2023Lya_escape_JWST,Tang2024LAE_JWST_z56,Lin2024_lya_escape_Ha_emitters,Witstok2024lyaz>8_diff_fesc,nakane2024high_z_LAEs} are starting to build larger sample sizes, but they are not yet at the quantity of LAEs and UV-selected galaxies used for the results in this work. In particular, intrinsically faint galaxies have poor statistics, although this can be improved with the continuing observations of lensing clusters by JWST/NIRCam and JWST/NIRSpec.\\
\indent On the other hand, it is possible, with further JWST observations, photometry, and spectroscopy in order to improve the \flya determinations for individual galaxies. The Balmer decrement provides accurate estimations of dust attenuation along with SED fitting utilising a greater number of photometric bands and a more extended coverage. Studying these systems in more detail will reveal the range of different escape scenarios leading to the varied \flya values that we have observed. This can, in turn, lead to better understanding of galaxy evolution around the epoch of reionisation and how LAEs affect this process. 


\begin{acknowledgements}
This work is done based on observations made with ESO Telescopes
at the La Silla Paranal Observatory under programme IDs 060.A-9345, 092.A-0472,
094.A-0115, 095.A-0181, 096.A-0710, 097.A0269, 100.A-0249, and
294.A-5032. Also based on observations obtained with the NASA/ESA \textit{Hubble}
Space Telescope, retrieved from the Mikulski Archive for Space Telescopes
(MAST) at the Space Telescope Science Institute (STScI). STScI is operated by
the Association of Universities for Research in Astronomy, Inc. under NASA
contract NAS 5-26555. 
All plots in this paper were created using Matplotlib (Hunter 2007). 
Part of this work was supported by the French CNRS, the Aix-Marseille University, the French Programme National de Cosmologie et Galaxies (PNCG) of CNRS/INSU with INP and IN2P3, co-funded by CEA and CNES.
This work also received support from the French government under the France 2030 investment plan, as part of the Excellence Initiative of Aix-Marseille University - A*MIDEX (AMX-19-IET-008 - IPhU).
Financial support from the World Laboratory, the Odon Vallet Foundation and VNSC is gratefully acknowledged. Tran Thi Thai was funded by Vingroup JSC and supported by the Master, PhD Scholarship Programme of Vingroup Innovation Foundation (VINIF), Institute of Big Data, code VINIF.2023.TS.108. This research was funded by Vingroup Innovation Foundation under project code VINIF.2023.DA.057.        
\end{acknowledgements}

\bibliographystyle{aa} 
\bibliography{aa51432_24corr} 

\begin{appendix} 
\section{LAE luminosity density and its star-forming rate density}

\setcounter{table}{0}
\renewcommand{\thetable}{A\arabic{table}}
 
\begin{table}[!h]

\caption{LAE luminosity density and its SFRD.}  

\begin{tabular}{ m{2cm}  m{3cm} m{1cm} m{1cm} m{1.2cm} m{2cm} m{2cm} m{2.3cm}}
\hline
\hline
        \smallskip
        Redshift & Reference & $\alpha$ & log $L*$ & log $0.0006L*$ &$\rho_{Ly\alpha}$ ($10^{39}$)& LAE SFRD ($10^{-3}$) & $f_{escp}\, [\%]$\\
        \hline
        \noalign{\smallskip}
        
       $z=3.1$ & \cite{Ouchi2008LAE_LF} &$-1.5$ (fix)& 42.76 & 39.54 & $9.21\pm2.89$ & $8.37\pm2.63$ & $5.21\pm1.85$
       \\
        \noalign{\smallskip}
        \hline
        \noalign{\smallskip}
     $z=3.7$ & \cite{Ouchi2008LAE_LF}& $-1.5$(fix) & 43.0 & 39.78 & $5.99\pm2.06$ & $5.44\pm1.87$ & $3.66\pm1.44$
     \\
        \noalign{\smallskip}
        \hline
        \noalign{\smallskip}
        $z=5.7$ & \cite{Ouchi2008LAE_LF} & $-1.5$ (fix) & 42.83& 39.61& $9.04\pm6.11$ & $8.21\pm5.55$ & $15.06\pm10.39$
        \\
        \noalign{\smallskip}
        \hline
        \noalign{\smallskip}
        $z=1.95-3$ & \cite{cassata2011vimosLAELF} & $-1.6$ & 42.74 & 39.52 & $7.45\pm3.15$ & $6.76\pm2.86$ & $3.57\pm1.63$
        \\
        \noalign{\smallskip}
        \hline
        \noalign{\smallskip}
       $z=3-4.55$& \cite{cassata2011vimosLAELF}& $-1.78$& 42.83& 39.61& $7.86\pm2.60$&$7.14\pm2.36$ & $4.86\pm1.87$
       \\
        \noalign{\smallskip}
        \hline
        \noalign{\smallskip}
        $z=4.55-6.6$ & \cite{cassata2011vimosLAELF}& $-1.69$ (fix) & $43.0$ & 39.78 & $12.42\pm3.10$ & $11.28\pm2.82$ & $18.5\pm5.4$
        \\
        \noalign{\smallskip}
        \hline
        \noalign{\smallskip}
        $z=1.9-3.8$ & \cite{blanc2011}& $-1.7$ (fix) & 43.08& 39.86 & $6.96\pm5.17$ & $6.32\pm4.70$ & $6.42\pm2.95$
        \\
        \noalign{\smallskip}
        \hline
        \noalign{\smallskip}
$z=3.1\pm0.3$ &\cite{Sobral_2018}&$-1.63$ &42.77&39.55&$11.44\pm6.73$ & $10.39\pm6.11$ & $6.47\pm3.95$
 \\
        \noalign{\smallskip}
        \hline
        \noalign{\smallskip}
        $z=3.9\pm0.2$ &\cite{Sobral_2018}&$-2.26$&42.93& 39.71 & $40.56\pm42.59$ & $36.83\pm38.68$ & $25.42\pm27.2$
         \\
        \noalign{\smallskip}
        \hline
        \noalign{\smallskip}
        $z=4.7\pm0.1$ & \cite{Sobral_2018}&$-2.35$ & 43.28 & 40.06 & $36.78\pm44.43$ & $33.40\pm40.35$ & $34.46\pm42.14$
         \\
        \noalign{\smallskip}
        \hline
        \noalign{\smallskip}
        $z=5.4\pm0.4$ & \cite{Sobral_2018}& $-1.98$& 43.28 & 40.06&$17.78\pm11.75$ & $16.14\pm10.67$ & $25.12\pm17.07$
         \\
        \noalign{\smallskip}
        \hline
        \noalign{\smallskip}
        $z=3.5$ & \cite{GdlV2019LAELF} & $-1.58$ & 42.77& 39.55 & $9.80\pm4.85$ & $8.90\pm4.41$ & $5.83\pm3.08$
        \\
        \noalign{\smallskip}
        \hline
        \noalign{\smallskip}
        $z=4.5$ & \cite{GdlV2019LAELF}& $-1.72$ & 42.96&39.74& $9.29\pm4.86$ & $8.43\pm4.42$ & $7.66\pm4.29$
        \\
        \noalign{\smallskip}
        \hline
        \noalign{\smallskip}
        $z=6.0$ & \cite{GdlV2019LAELF}& $-1.87$ & $43.16$ & 39.94 & $9.52\pm5.65$ & $8.64\pm5.13$ & $15.85\pm9.66$
        \\
        \noalign{\smallskip}
        \hline
        \noalign{\smallskip}
        $z=3.5$ & \cite{Thai2023} & $-2.0$ & 42.87& 39.65& $33.34\pm15.19$ & $30.27\pm13.79$ & $19.82\pm9.73$
         \\
        \noalign{\smallskip}
        \hline
        \noalign{\smallskip}
        $z=3.5$ & \cite{Thai2023} (modified Schechter function) & $-2.03$ & 42.90& 39.68 & $27.69\pm11.24$ & $25.15\pm10.21$ & $16.47\pm7.33$
        \\
         \noalign{\smallskip}
        \hline
        \noalign{\smallskip}
        $z=4.5$ & \cite{Thai2023}& $-1.97$ & 42.97& 39.75 & $23.17\pm11.44$ & $21.04\pm10.39$ & $19.12\pm10.16$
        \\
        \noalign{\smallskip}
        \hline
        \noalign{\smallskip}
$z=4.5$ & \cite{Thai2023} (modified Schechter function)& $-2.06$ & 43.04 & 39.82 & $17.47\pm11.01$ & $15.9\pm10.0$ & $14.41\pm9.52$        
        \\
        \noalign{\smallskip}
        \hline
        \noalign{\smallskip}
        $z=6.0$ & \cite{Thai2023}& $-2.28$ & 43.03& 39.81 & $103.23\pm76.5$ & $93.74\pm69.47$ & $171.94\pm129.61$
        \\
        \noalign{\smallskip}
        \hline
        \noalign{\smallskip}
$z=6.0$ & \cite{Thai2023} (modified Schechter function)& $-2.67$ &43.26&40.04& $47.31\pm37.63$ & $42.96\pm34.17$ & $78.8\pm63.61$
        \\
        \hline
               
        \vspace{0.2cm}
        \end{tabular}

    \label{tab: LAE luminosity density}
    \vspace{-5mm}
    \begin{minipage}{0.95\textwidth }
    \vspace{0.1cm}
    \begin{tablenotes}
      \small
      \item \textbf{Notes.} The SFRD was calculated by integrating each LF to $0.0006L*$. The annotation `fix' denotes slopes that were fixed by the authors of the study in question. The units of $\rho_{Ly\alpha}$ and the LAE SFRD are $\mathrm{erg\, s^{-1} Mpc^{-3}}$ and $\mathrm{M_{\odot} yr^{-1} Mpc^{-3}}$ respectively and the units of $L*$ are $\mathrm{erg \, s^{-1}}$.
    \end{tablenotes}
    \end{minipage}
\end{table}
\end{appendix}

\end{document}